# Well-Behaved Model Transformations with Model Subtyping

Artur Boronat
Department of Informatics
University of Leicester
aboronat@le.ac.uk

## ABSTRACT

In model-driven engineering, models abstract the relevant features of software artefacts and model transformations act on them automating large tasks of the development process. It is, thus, crucially important to provide pragmatic, reliable methods to verify that model transformations guarantee the correctness of generated models in order to ensure the quality of the final end product. In this paper, we build on an object-oriented algebraic encoding of metamodels and models as defined in the standard Meta-Object Facility and in tools, such as the Eclipse Modeling Framework, to specify a domain-specific language for representing the action part of model transformations. We introduce the big-step operational structural semantics of this language and its type system, which includes a notion of polymorphic model subtyping, showing that well-typed model transformations are well behaved. That is, that *metamodel-conformant model transformations never go wrong*. Both the interpreter and the type system are implemented and available online.

## Keywords
Meta-Object Facility, reuse in model transformation, structural operational semantics, type theory.

## 1. INTRODUCTION

Model-driven software development (MDSD) initiatives, such as OMG's Model-Driven Architecture (MDA) or, more generally, model-driven engineering (MDE), aim at improving quality, productivity and cost-effectiveness in software development processes [50]. This is achieved both by representing models of software artefacts using a uniform exchange format and by enabling their systematic treatment using dedicated model transformation languages. The first part may be addressed by employing the Meta-Object Facility (MOF) standard, which offers a generic framework in which the abstract syntax of modelling languages can be defined [54]. The second part is usually addressed by means of dedicated model transformation languages that allow for the development of model compilers as model transformations within one or among several metamodels [53].

Demonstrating the reliability of such model transformations and of the generated software is a crucial step for MDE to succeed. Recent surveys [2, 4, 29] provide an outline of verification techniques applied to model transformations, ranging from lightweight approaches based on testing to automated and interactive theorem proving. Considering model transformations as models [21, 5], verification of model transformations can be performed at the model level, where reasoning is applied to specific model transformations, or at the metamodel level, where reasoning is applied to a class of model transformations specified by a model transformation system [59]. In this paper, we are interested in verifying meta-properties of model transformation languages and associated tooling in order to ensure the correctness of model transformations in general, while providing pragmatic tools support for engineers. Specifically, we reason about meta-properties of interpreters and of type systems for the model actions of a model transformation language.

Considering properties of the generated models, the correctness of model transformations can be studied from two broad points of view: syntactic correctness and semantic correctness. A model transformation is considered syntactically correct [30, 18] if all possible input models are transformed into models that conforms to their MOF metamodel, either by using model types or by considering additional constraints (usually OCL or graph constraints). When models are augmented with behavioural semantics, semantic correctness of model transformations [32] implies some sort of behaviour preservation using (bi-)simulation techniques. Since we are focussing on MOF, which does not allow for the specification of behaviour semantics in metamodels, we equate the term correctness to its syntactic variant in the rest of the paper.

Regarding the effects of model transformations on models, we decompose the notion of EMF model graph [8] to identify two sources of errors depending on whether they can be considered (checked or avoided) at run time or earlier, statically at compilation time: errors regarding the structure of a model when references and containments are modified (e.g. violation of composition semantics, generation of dangling references, consistency of opposite references) and errors with respect to the model types declared in the metamodel. Model transformations that do not cause errors of the first kind are called safe. When, in addition, they do not cause errors of the second kind, they are called well behaved and a type system is required to characterize them.

The goal of a type system [19, 47] is to provide a quick, short feedback loop for programmers in order to find runtime errors while writing a program. The design and implementation of a type system can be fairly involved but, once the pieces of the machine are clicked together appropriately, a type system provides a lightweight tool for ensuring the correctness of a program. The correctness of a type system for model transformations can be analyzed from two points of view [46]: a type system is syntactically sound when it verifies that the program belongs to the model transformation language as defined by the type system; and a type system is semantically sound when, in addition, it ensures that well-typed model transformations are well behaved.

In this work, we propose the semantics of a model action language for implementing model transformations that is safe, and a type system for MOF model types and for the action language. The type system is based on an algorithm for structural type inference that supports the notion of model subtyping (subsumption). The type system is shown to be sound with respect to the semantics of the action language, ensuring that well-typed model transformations are well behaved.

The overall contribution of the paper is, thus, a strongly typed language for the action part of model transformations that can be used to ensure correct model transformations. At the moment, most of the model transformation languages available are weakly typed, apart from those directly implementing graph transformation theory (in particular, using typed attributed graphs with inheritance). More specific contributions in this paper are enumerated as follows:

- a characterization of the notions of safe model transformations and of well-behaved model transformations with respect to the notion of EMF model graph;

- a specification of the big-step structural operational semantics of a language of focussed model actions (FMA), whose programs are safe;

- a specification of a type system for models based on structural type inference that does not require code generation for the metamodel, providing support both for implicit, duck typing[1] and for model subtyping in model transformations;

- a specification of a type system for FMA programs that is semantically sound, guaranteeing that well-typed FMA programs are well behaved.

The results are developed using standard notations and techniques normally employed in the design of programming languages and of their tool support, independently of concrete implementations. Therefore, they can be used as a reference for implementing model transformation languages. In addition, a reference implementation of the interpreter and of the type system are available online[2], providing further help in that direction, e.g. by enabling testing of other implementations or by providing building blocks for other model transformation languages. For example, we have implemented JSON upsert-delete[3], a model transformation language for (untyped) models that have been JSONized as JSON document graphs, using a variant of the FMA interpreter.

The structure of the paper is as follows. In section 2, we discuss current approaches to reuse in model transformations and we present a novel type system for models based on structural type inference. In section 3, we discuss notions of correctness for model transformations based on the notion of EMF model graph, and we present the syntax, big-step structural operational semantics of a language of focussed model actions, and a type system for FMA programs. FMA programs are shown to be safe and well-typed FMA programs are shown to be well behaved. In section 4, we discuss related work regarding the correctness of model transformations from a program correctness point of view and from a type-theoretic point of view.

---
[1] https://en.wikipedia.org/wiki/Duck_typing
[2] http://arturboronat.info/fma
[3] http://arturboronat.info/jsonud

## 2. MODEL TYPES AND REUSE

A common way to represent models (in the MOF 2.0 sense) is to treat them as a flat configuration of objects, as in classical object-oriented systems [45, 3], with (potentially) bidirectional relationships [55]. MOF is, however, a specification, and specific realizations, such as EMF, make it clear that there are other features, such as composite association ends (the opposite of containments in EMF parlance), that induce more complex structure in model representation. In the research community, this is evidenced by a number of approaches that either specialize graph transformation languages to deal with model transformations, e.g. [7, 6, 8], or explore alternative representations where the hierarchical nature of models is exploited, e.g. [14].

In this section, we present a type system based on structural type inference that provides a notion of model type based on subsumption, which enables us to study a sound type system for model transformations. In this paper, we focus on the kernel of MOF without considering multiplicity and OCL constraints. Such a compromise is not unusual in formal approaches to model transformations, such as those based on graph transformation, in order to enable the use of automated decision procedures in MDE - in this case, type checking.

In the following subsections, we start by discussing approaches for model typing, focussing on accepted notions of model subtyping for MOF metamodels and for graph transformation theory. Some considerations with an example that motivate the need for a new notion of model subtyping are then discussed. A syntactic representation for models, a notion of model type and a notion of model subtyping are introduced by providing a type system.

### 2.1 Model Subtyping

In the recent literature on model transformation, there is an emerging interest in formalizing mechanisms for reuse of model transformations [20, 43, 37]. Approaches to reuse transformation logic (captured as transformation intent in [44]) involve mechanisms to facilitate its application in different contexts (by means of typing) or by extending the logic itself (by means of transformation rule extension). We are going to focus on the first one in order to study when it is safe to reuse a model transformation.

Regarding typing, working with models in MDE processes can be done at two levels of abstraction: a white-box approach (called intra-resource in [60]) where model internals are exposed so that side effects can be analyzed as model transformations act upon models; and a more coarse (black-box) approach (called inter-resource in [60]) where models are treated as first-order citizens in model management scenarios [13]. The former is a necessary building block to enable a sound basis for the latter, i.e. by building model management systems based on safe model transformations.

Representative works of the latter approach provide type systems for model management languages. Vignaga et al. [60] studied a safe language for defining higher-order trans-formations (HOTs) based on dependent type calculus, where model types are considered as scalar types for variables and type checking does not inspect the internals of given models. Chechik et al. [20] discuss a more general form of reuse that involves the use of model transformations as conversion operators, which can play an adaptation role. Conversion operators must obey a number of laws in order to induce a coercive model type system, where the specific side effects of transformations are abstracted away.

Further refining this classification around the notion of intra-resource typing, we find approaches that build on a notion of model type, typically considering model subtyping or metamodel adaptation.

Model subtyping can be dealt with as a subsumption relation or with model type matching, by generalizing the homologous notions in OO programming languages [3]. Steel et al. [55] proposed a type system, implemented in Kermeta [40], using a notion of model type matching, where a model type M' matches a model type M, denoted M' <# M, iff for each class C in M, there is a class C in M' such that the signature of its operations is preserved[4]. Guy et al. improved this notion of subtyping as isomorphic model subtyping in [37] and introduced non-isomorphic model subtyping for enabling model adaptation by means of renaming maps. The language of model transformations used in [55, 37] refers to in-place procedures that may not be functional, thus not guaranteeing compositionality, and it does not ensure the semantic soundness of the type system. That is, the type system can be used to statically find bugs in a model transformation but it does not guarantee that well-typed transformations behave correctly (with respect to the properties encoded in the type system). Guy et al. [37] also discuss the notion of partial and total subtyping in order to facilitate reuse of model transformations in practical scenarios. Partial model subtyping aims at enabling the safe reuse of a model transformation even if only the part of the model type that is used in the model transformation is present. In this context, Sen et al. [52, 51] conceptualized this situation as the notion of effective model type of a transformation: the minimal subset of the elements of the input metamodel that is used in the transformation.

Model transformations based on graph transformation theory rely on the theory of typed attributed graphs with type node inheritance [31, 25]. Typing checking in this theory is achieved by constructing a graph morphism between a graph (the model) and the type graph (the metamodel) that preserves the structure of the graph. Model subtyping is supported in graph transformations by means of the notion of abstract production rule where nodes in a graph pattern in the rule may correspond to abstract nodes (similarly to an abstract superclass). From a theoretical point of view, given a graph and an abstract production rule that can be applied to it[5], it has been shown that a unique concrete production rule can be constructed so that the effects of the transformation on the graph are equivalent to the application of abstract production rule directly on the graph. This means that the usual theory for typed attributed graph transformation can be applied for graph grammars with abstract production rules, with a notion of object subtyping.

A representative approach to metamodel adaptation draws on the definition of generic model transformations using concepts [26], which involves the definition of explicit bindings between a concept (supertype) and a metamodel (subtype). Model transformations (model templates) are defined with respect to the concept and later executed over instances of a concrete metamodel by generating concrete model transformations using higher-order transformations (HOTs) [22]. Thus, this approach is useful for handling adaptation of model transformations to different contexts in a non-intrusive way. A notion of algebraic adapters was presented in [27] for facilitating adaptation both of a model transformation, as in [22], and of a meta-model, in the sense of [37]. Thus, model transformations can be defined around the notion of concepts and these can be reused in a wide myriad of contexts, even when the target metamodel (and its models) need to be augmented with information. Legacy model transformations can also be reverse engineered as model transformations defined over concepts [23].

## 2.2 Example

In this subsection, we discuss the notion of model type matching and typing in graph transformation theory with an example. In the example, we are using the metamodel for defining graphs and the metamodel for defining state machines shown in Figure 1. The model types described by both metamodels are structurally similar in that they both describe languages of graphs.

*Considerations about model type matching [37, 55].*

First, they consider a model type and a metamodel as interchangeable concepts and, in fact, their type checking algorithms work over the representation of a model type in the form of a MOF metamodel (or Ecore model in EMF). In our approach we prefer to keep the notion of model type and of its MOF representation separate [12, 11].

Second, they motivate the need for a model subtyping mechanism other than subsumption because they consider properties as pairs of generator-mutator methods, which requires invariance of property types in order not to violate Liskov's substitution principle. This feature makes sense in the context of their approach, where meta-classes in meta-models contain operation signatures whose behaviour is defined in Kermeta. However, if metamodels are regarded as purely syntactic devices and behaviour is defined by means of external model transformations (e.g. by model compilers or by simulators) there is no need for declaring operations in metamodels. This observation opens the door for exploring subsumption for model subtyping.

Third, the notion of model type enables the definition of models with different roots even when those objects are not instance of root meta-classes. For example, when there is a unidirectional composition between a composite meta-class A and a component meta-class B. In such cases, a model formed by two root objects, one of type A and one of type B would be well typed. A problematic situation emerges when the model type is used as the type of the parameter of a model management operation, where only objects of root

---
[4] Noting that properties are encoded as pairs of generator-mutator methods in their approach.

[5] That is, there is a consistent match from the left-hand side graph of the rule to the graph, meaning that: the match preserves the so-called gluing condition in order to ensure the graph structure; it preserves the typing − considering subtyping declared as type node inheritance − of the pattern; and it satisfies the negative application conditions of the rule.

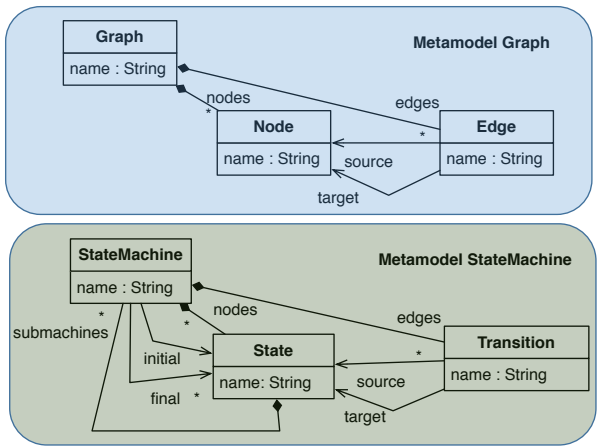

Figure 1: Two metamodels.

meta-classes are expected as arguments.

Fourth, the type system proposed in [55] relies on class names to match object types, which implies that the model type for state machines would not be considered a subtype of the model type for graphs.

*Considerations about typed attributed graphs.*

First, for a graph to be considered well typed one needs to show that there is an explicit total graph homomorphism between the graph and the type graph as explained above. That is, typing is defined using an ontological approach [42], where the *is instance of* relation is explicitly declared with a set of references from objects to their types. This means that if the type graphs in a typed graph transformation system are removed, the graph to be modified becomes a rather flat structure without ability to distinguish among node elements and among edge elements. The type graphs representing the metamodels in Figure 1 denote different types of graphs that are not related unless adaptation mechanisms are used explicitly, e.g. by using concepts. We prefer to consider a model representation where typing information is already embedded in the structure of the model following a linguistic approach [42]. This allows us to reuse the theory behind programming language design for building MDE tools for automating model transformation and model management tasks.

On the other hand, the encoding of side effects of model transformations in a rule of a graph grammar are dependent on the matches found for the left-hand side pattern of the rule. This complicates the operationalization of declarative model transformations if one is not interested in a language with pattern matching, e.g. the desired model transformation language may provide support for queries. We prefer to decouple the definition of side effects from a pattern matching mechanism in order to offer language support both for declarative and for deterministic model transformation languages. That is, we prefer to develop a DSL that can be used to implement any model transformation language, including rule based ones.

## 2.3 Model Representation

In this section, we describe how to represent models in terms of structured objects, i.e. aggregates in domain-driven development terminology [33] or composite objects in UML [9], using an object-oriented notation. In the UML, the notion of model is determined by a class from the metamodel designated as root class. A model is an instance of that class together with all its containments. MOF 2.0 introduced the notion of extent, allowing the representation of models with configurations of objects instantiated from different packages, consequently a model may consist of several root objects. We follow the approach taken in [55] and consider models as (ordered) sets of structured objects.

Given the domains $\mathcal{D}_b$ for the base type names $b \in \mathcal{B}$, the countable set $\mathcal{O}$ of object identifiers $o$, and the countable set $\mathcal{P}$ of property names $p$, the syntax for representing models is as follows:

$$\begin{aligned}
\textbf{Term} \ni t &::= v \mid is \mid os & \text{(terms)} \\
\textbf{OidSet} \ni is &::= o \mid is\ is \mid \emptyset & \\
& & \text{(ordered sets of object identifiers)} \\
\textbf{ObjSet} \ni os &::= \rho \mid os\ os \mid \emptyset & \text{(ordered sets of objects)} \\
\textbf{Object} \ni \rho &::= <o \mid ps> & \text{(objects)} \\
\textbf{PS} \ni ps &::= p:t, ps \mid \emptyset & \text{(sets of object properties)}
\end{aligned}$$

where we are using $v \in \mathcal{D}_b$ as a meta-variables for base type values. We use the following auxiliary operations $\_{}_o$ : $\textbf{Object} \to \mathcal{O}$, $\_{}_{ps}$ : $\textbf{Object} \to \textbf{PS}$ to project the components of an object $\rho \in \textbf{Object}$.

*Definition 1.* (Structured Model) Given a countable set $\mathcal{C}$ of class names, a structured model $M$ is a pair $(M, \Pi)$ where $M$ is an ordered set $os$ of objects and $\Pi$ is a typing environment $\mathcal{O} \to \mathcal{C}$ mapping the identifiers of all the objects in $M$ to a class name.

Object sets $os$ representing structured models do not carry type information explicitly by using type names, as in [12]. Typing information is implicitly encoded in the structure of the term representing the model and explicitly in the typing environment $\Pi$. In our approach, we consider graphs as a special case of structured models, of depth 0, although graphs can occur at any depth level in the model.

## 2.4 Model Type Representation

Models in software engineering have a dual interpretation, namely as "a related collection of instances of meta-objects, representing (describing or prescribing) an information system, or parts thereof, such as a software product or as semantics", or as "a semantically closed abstraction of a system or a complete description of a system from a particular perspective" [1]. That is, as syntax or as semantics [38]. Since MOF metamodels are also models, we apply this dual interpretation to them and differentiate a MOF metamodel from a MOF model type by saying that a metamodel denotes a unique model type [12, 11].

Our model types are used as classifiers of structured objects whose operations are defined for models, even if they act on the internals of the model representation. For example, a model transformation is defined as an FMA sequence of statements that may update property values and alter the structure of the model. However, meta-classes do not contain operation signatures.

Model types are inferred from a set of objects representing a model, providing support for implicit typing and for duck typing - required for considering the state machine model type as a subtype of the graph model type in Figure 1.

*Definition 2.* (Syntax for types) Given the finite sets $\mathcal{B}$ of base type names $b$, $\mathcal{C}$ of class names $c$, $\mathcal{P}$ of property names $p$, the set $\tau$ of types over $\mathcal{B}$, $\mathcal{C}$ and $\mathcal{P}$, is defined as follows

$$\textbf{Type} \ni \tau ::= \alpha \,|\, \varsigma$$
$$\textbf{ObjectType} \ni \varsigma ::= p : \alpha, \varsigma \,|\, \emptyset \,|\, \bot$$
$$\textbf{Scalar} \ni \alpha ::= b \,|\, ref\ c? \,|\, () \,|\, \bullet$$
$$\textbf{MaybeClassName} \ni c? ::= c \,|\, \texttt{Any}$$

where $b$ is a type in $\mathcal{B}$, $ref\ c$ corresponds to a reference type, $()$ and $\bullet$ are unit types used for the type system of the model transformation language, $\emptyset$ denotes the empty set, $\bot$ denotes the bottom type as the least informative object type and $\texttt{Any}$ denotes its name.

*Remark 1.* $p_1 : \alpha_1, \ldots, p_n : \alpha_n$ corresponds to a set of structural features, where attributes are defined as properties of the form $p : b$, references pointing to a class $c$ are defined as properties of the form $p : ref\ c$, and containments pointing to a class $c$ are defined as properties of the form $p : c$.

*Definition 3.* (Model Type) Given finite sets $\mathcal{B}$ of base type names $b$[6], $\mathcal{C}$ of class names $c$, and $\mathcal{P}$ of property names, a model type $\mathcal{M}$ is defined as a tuple $(\mathcal{B}, \mathcal{C}, \mathcal{P}, cl, df, oe, sr, r)$ where:

- $cl$ is an injective function assigning to each class name $c$ a corresponding object type $\varsigma$ of the form $p_1 : \alpha_1, \ldots, p_n : \alpha_n$ that specifies the structure of the objects of class $c$.

- $df$ is a function, which receives a class name $c$ and an object identifier $o$ as arguments, assigning to each class name $c$ a default property set $ps$, where properties are initialized to their default values (the opposite reference to the containment, when it exists, is initialized with a reference to the identifier $o$ of the container object).

- $oe$ is a bijective function $\texttt{BRE} \to \texttt{BRE}$ where $\texttt{BRE}$ is a set of both reference ends − of the form $\texttt{bRE}(c,p)$ − participating in bidirectional associations, and containment ends − of the form $\texttt{bCE}(c,p)$ − participating in bidirectional compositions. $oe$ cannot map containment ends to containment ends. In addition, we have that $(\texttt{BRE}, oe)$ forms a group [35], ensuring that a bidirectional reference or composition is formed by only two reference ends or by a reference end and by a containment end.

- $sr$ is the reflexive transitive closure of the relation formed by pairs of the form $c\texttt{<:}c'$, where $c, c' \in \mathcal{C}$ are class names corresponding to an object subtype and to an object supertype, respectively.

- $r \in \mathcal{C}$ is the name that designates the root meta-class in the metamodel.

## 2.5 Type System for Models

In this section, we present the notion of object subtyping (subsumption) and how it is generalized for a model type.

---

[6] In the current version of the tool, we only consider String and Integer.

We build on these notions to define the type system for models that provides type inference for models.

Model subtyping emerges from object subtyping and it is implicitly defined when a metamodel is given. That is, the user does not have to deal with dedicated syntactic constructs to define the relation upfront. Our type system checks whether subtyping is preserved.

*Definition 4.* Object Subtyping (Subsumption)

Given a model type $\mathcal{M}$, the object subtyping relation $<: \subseteq \textbf{ObjectType} \times \textbf{ObjectType}$ is defined by the pairs of object types $(\varsigma', \varsigma)$ that satisfy the following rules when $\varsigma \neq \emptyset$:

$$\forall (p : b) \in \varsigma \Rightarrow \exists (p : b') \in \varsigma' \text{ such that } b = b'$$
$$\forall (p : c) \in \varsigma \Rightarrow (\exists (p : c') \in \varsigma' \text{ such that } cl(c') <: cl(c))$$
$$\vee \exists (p : \texttt{Any}) \in \varsigma'$$
$$\forall (p : ref\ c) \in \varsigma \Rightarrow (\exists (p : ref\ c') \in \varsigma' \text{ such that } cl(c') <: cl(c))$$
$$\vee \exists (p : ref\ \texttt{Any}) \in \varsigma'$$

Moreover, $\emptyset$ subsumes any object type $\varsigma$ and any object type $\varsigma$ subsumes $\bot$.

We generalize the notion object subtyping to the notion of model subtyping through the object type of the root objects of a model as follows:

*Definition 5.* Model Subtyping (Subsumption). A model type $\mathcal{M}$ is said to be a subtype of a model type $\mathcal{M}'$, denoted by $\mathcal{M} <: \mathcal{M}'$, if and only if $cl(\mathcal{M}_r) <: cl'(\mathcal{M}'_r)$.

When a model consists of several root objects, as explained in [55], whose type is defined in different metamodels, the root objects must be related via subsumption. For example, when extending several metamodels that provide different root meta-class names, the extending metamodel must include a root meta-class as superclass of each of the root meta-classes of the extended packages. The reason for this is that our type system considers that collections of objects are homogeneous (albeit polymorphic via object subtyping).

The type system for models is given in Table 1 and in Table 2. Typing judgements are of the form $\Gamma\,|\,\Pi \vdash t : \tau$ stating that the term $t$ has type $\tau$ given the environments $\Pi$ and $\Gamma$, where $\Gamma : \textbf{Var} \rightharpoonup \textbf{Type}$ is a typing environment for variables and $\Pi : \mathcal{O} \rightharpoonup \mathcal{C}$ is a typing environment for object identifiers. Both typing environments are considered partial injective maps.

The type system consists of two types of inference rules, with premises (P) and conclusions (C), using the notation $\frac{P}{C}$: axioms, where there is no premise, and inductive rules that explain how conclusions are inferred from premises subject to the satisfaction of side conditions, usually co-located together with the premises P.

The axioms in our type system are as follows: (T-BASE) determines the type of a value, (T-REF) determines the type of a reference based on the typing environment for object identifiers, (T-REFANY) determines the type of an undefined reference, (T-OBJECTANY) determines the type of an undefined containment, (T-PROPEMPTY) determines the type of an empty set of properties.

The typing rules are defined as follows: (T-OBJ) finds the type of an object based on the type inferred from its

$$\frac{a \in D_b}{\Gamma \mid \Pi \vdash a : b} \quad \text{(T-Base)}$$

$$\frac{\Pi(o) = c}{\Gamma \mid \Pi \vdash o : \mathit{ref}\ c} \quad \text{(T-Ref)}$$

$$\Gamma \mid \Pi \vdash \mathit{ref}(\emptyset) : \mathit{ref}\ \bot \quad \text{(T-RefAny)}$$

$$\frac{\Gamma \mid \Pi \vdash o : \mathit{ref}\ c \quad \Gamma \mid \Pi \vdash \mathit{ref}(is) : \mathit{ref}\ c' \quad cl(c) <: cl(c')}{\Gamma \mid \Pi \vdash \mathit{ref}(o\ is) : \mathit{ref}\ c'} \quad \text{(T-Ref1)}$$

$$\frac{\Gamma \mid \Pi \vdash o : \mathit{ref}\ c \quad \Gamma \mid \Pi \vdash \mathit{ref}(is) : \mathit{ref}\ c' \quad cl(c') <: cl(c)}{\Gamma \mid \Pi \vdash \mathit{ref}(o\ is) : \mathit{ref}\ c} \quad \text{(T-Ref2)}$$

$$\frac{\Gamma \mid \Pi \vdash o : \mathit{ref}\ c \quad \Gamma \mid \Pi \vdash \mathit{ref}(is) : \mathit{ref}\ c' \quad cl(c') \not<: cl(c) \quad cl(c) \not<: cl(c') \quad c'' = \wedge(c, c', \mathcal{M}_{sr})}{\Gamma \mid \Pi \vdash \mathit{ref}(o\ is) : \mathit{ref}\ c''} \quad \text{(T-Ref3)}$$

**Table 1: Typing rules for models.**

$$\frac{\Pi(o) = c \quad \Gamma \mid \Pi \vdash (p_1 = u_1, \dots, p_n = u_n) : \varsigma \quad \varsigma =: cl(c)}{\Gamma \mid \Pi \vdash \texttt{<}o \mid p_1 = u_1, \dots, p_n = u_n\texttt{>} : c} \quad \text{(T-Obj)}$$

$$\Gamma \mid \Pi \vdash \emptyset : \bot \quad \text{(T-ObjAny)}$$

$$\frac{\Gamma \mid \Pi \vdash \rho : c \quad \Gamma \mid \Pi \vdash os : c' \quad cl(c) <: cl(c')}{\Gamma \mid \Pi \vdash \rho\ os : c'} \quad \text{(T-Obj1)}$$

$$\frac{\Gamma \mid \Pi \vdash \rho : c \quad \Gamma \mid \Pi \vdash os : c' \quad cl(c') <: cl(c)}{\Gamma \mid \Pi \vdash \rho\ os : c} \quad \text{(T-Obj2)}$$

$$\frac{\Gamma \mid \Pi \vdash \rho : c \quad \Gamma \mid \Pi \vdash os : c' \quad cl(c') \not<: cl(c) \quad cl(c) \not<: cl(c') \quad c'' = \wedge(c, c', \mathcal{M}_{sr})}{\Gamma \mid \Pi \vdash \rho\ os : c''} \quad \text{(T-Obj3)}$$

$$\Gamma \mid \Pi \vdash \texttt{none} : \emptyset \quad \text{(T-PropEmpty)}$$

$$\frac{\Gamma \mid \Pi \vdash t_1 : \alpha_1 \quad \Gamma \mid \Pi \vdash (p_2 = t_2, \dots, p_n = t_n) : \varsigma}{\Gamma \mid \Pi \vdash (p_1 = t_1, \dots, p_n = t_n) : (p_1 : \alpha, \varsigma)} \quad \text{(T-Prop)}$$

**Table 2: Typing rules for models.**

property set by using the operator $=:$ (where $\varsigma =: \varsigma'$ is defined as $\varsigma|_{names(\varsigma')} = \varsigma \wedge \varsigma <: \varsigma'$, that is it ensures that the inferred object type has the same properties as in the object type definition considering that some inferred property types may be $\bot$ or $\mathit{ref}\ \bot$); (T-Prop) infers an object type from a property set; (T-Ref1), (T-Ref2) and (T-Ref3) infer the type of a collection of references by selecting the most general type of the objects referenced that is the least informative one at the same time by means of the meet operator $\wedge$ (where the expression $\wedge(c, c', \mathcal{M}_{sr})$ obtains the class name of the least informative type that is supertype of both $\mathcal{M}_{cl(c)}$ and $\mathcal{M}_{cl(c')}$); similarly, (T-Obj1), (T-Obj2) and (T-Obj3) infer the type of a collection of objects by selecting the most general type of the objects referenced that is the least informative one at the same time. To clarify the purpose of the last six rules we take an example from the `Ecore` meta-modelling language of the Eclipse Modeling Framework (EMF). In any EMF metamodel, `EObject` corresponds the most general object type although it is not very informative when we compare it with other concrete metaclasses in a meta-model. However, `EObject` is still a useful type, as we are forced to work with it via the reflective API when other meta-classes are not present.

In our model representation, references and containment properties are bound to (ordered) sets. That is, more formally, the domains for references and containments are powersets and not just sets so that a value is a set of references or of objects. At the moment, we are only considering string, integer and boolean as base types.

Given finite sets $\mathcal{O}$ of object identifiers $o$ and $\mathcal{C}$ of class names $c$, the reference typing $\Pi : \mathcal{O} \to \mathcal{C}$ assigns class names to each of the object identifiers used in a model instance so that $o$ has type $\mathit{ref}\ c$ if and only if $\Pi(o) = c$ (as considered in rule T-Ref) in Tables 1 and 2).

*Definition 6.* Metamodel conformance. Given a metamodel, whose model type is $\mathcal{M} = (\mathcal{B}, \mathcal{C}, \mathcal{P}, cl, df, oe, r)$, we say that the structured model $M = (os, \Pi)$ conforms to the metamodel, denoted $M : \mathcal{M}$, iff $\emptyset \mid \Pi \vdash os : c$ and $\mathcal{M}_{cl}(c) <: \mathcal{M}_{cl}(\mathcal{M}_r)$.

As argued in [55], a mechanism for structural model subtyping is dependent on the representation of models. However, we have used MOF as the core backbone behind this representation and the type system is implemented for EMF metamodels. In this way, we guarantee that our type system will work for MOF-compliant tools. In addition, our notion of model subtyping provides a mechanism for reuse of models and model transformations defined in other tools as far as they are MOF compliant without using adaptation mechanisms and without having to generate extra code. Thus, our model type system provides support for duck typing and for implicit typing, while facilitating a flexible approach to reuse of model transformations.

## 3. DSL OF FOCUSSED MODEL ACTIONS

In this section, we present the specification of a DSL of Focussed Model Actions (FMA), modelling the typical side effects found in a model transformation: object creation and destruction; setting and unsetting structural features, which can be attribute values, cross-references or containments.

The main goal behind the design of this DSL is to characterize model transformations that are always well behaved. To explain what we mean by well-behaved transformation, we introduce the notion of valid model, capturing the structure of a model, and we adapt the notion of EMF consistent model [8], expressing when a valid model also conforms to its metamodel, as explained in Definition 6.

Based on the notion of valid model, we show that FMA is a safe language, in the sense that given a FMA model transformation program and a valid input model, the interpreter will always produce a valid output model, or else it will produce an anticipated error. To show this result, we introduce the FMA's syntax with the example of pull up attribute refactoring [34] and we explain the big-step structural operational semantics [41, 48] of the DSL, including the main components required for defining the semantics and two evaluation relations.

Based on the notion of consistent model, we show that well-typed FMA model transformation programs are well behaved, in the sense that the execution of a well-typed model transformation program always computes a consistent model. To show this result, we specify the type system for FMA programs, including its syntax and the typing relation in subsection 3.4. The main result of the work is the consistency theorem of subsection 3.5, showing that FMA is type sound and, consequently, that well-typed FMA model transformation programs are well behaved. More informally, that metamodel-conformant model transformations never go wrong.

## 3.1 Valid and Consistent Models

In this section, we explain the intuition behind the notions of valid model and of consistent model used to characterize when a model transformation is safe in section 3.4 or well-behaved in section 3.5, respectively. The notion of consistent model decouples the defining conditions of the formal notion of EMF model graph given in [8, Definition 3], which is also aligned with what a model is according to [55, section 2]. This redefinition helps us to classify the defining conditions of an EMF-model graph under two separate categories, which characterize different types of errors as explained below. This notion refines the notion of model type value presented in [12], where the representation was based on flat configurations of objects analogous modelling algebraic graph structures.

Given that we are starting from a formal notion based on graph theory, we will start by analysing its properties from a graph-oriented perspective. Then, we distil those properties that are inherent to the representation chosen in our approach, given in subsection 2.3, and we provide a formal definition based on our representation. According to [8, Definition 3], a model is considered valid iff it has the following structural properties:

1. objects and references form a graph, where referential integrity is key;

2. each object has at most one container;

3. there are not containment cycles;

4. opposite references are consistent;

5. there are no parallel edges.

Conditions 2, 3 and 5 are guaranteed by the representation described in subsection 2.3. The last condition is optional if bags and sequences are used when a metamodel is declared. However, we use ordered sets for references and for containments in our work, ensuring this condition as well. We use the remaining two properties to characterize a valid structured model as follows.

*Definition 7.* Valid structured model. Given a model type $\mathcal{M}$, a structured model $M$ is valid, denoted $M : \mathcal{M}_{pre}$, iff:

1. for any reference $p = ref(o\ is)$, there is an object in the model with object identifier $o$[7].

2. for any two opposite references $p$ in meta-class $c$ and $p'$ in meta-class $c'$, $bRE(c',p') = \mathcal{M}_{oe}(bRE(c,p))$, including the case where the opposite reference is a containment $bRE(c',p') = \mathcal{M}_{oe}(bCE(c,p))$ or $bCE(c',p') =$ $\mathcal{M}_{oe}(bRE(c,p))$, if both an object with identifier $o$ points to an object with identifier $o'$ through reference $p$ and the object with identifier $o'$ points to the object with identifier $o$ through reference $p'$.

---
[7]In the SOS specification, this statement is formalized by using object locations in subsection 3.3.1.

*Definition 8.* Consistent structured model. Given a model type $\mathcal{M}$, a structured model $M$ is consistent, denoted $M : \mathcal{M}$, iff both it is valid and it conforms to its metamodel, i.e. iff

$$M : \mathcal{M}_{pre} \wedge M : \mathcal{M}$$

In FMA's SOS semantics, we can distinguish two types of run-time errors: trapped errors, which cause the execution to stop, and untrapped errors, which stay hidden when they occur, potentially causing damage at a later point at run time. Trapped errors correspond to the violation of a defining condition of the notion of valid model, and are summarized in [10]. Untrapped errors correspond to errors that prevent the model from conforming to its metamodel. Errors of this kind are detected by the type system explained in section 2.

Therefore, trapped errors cause an invalid model and untrapped errors cause an inconsistent model, which could also be invalid. That is an invalid model is also inconsistent but an inconsistent model may be valid.

## 3.2 FMA Syntax

In this subsection, we provide the syntax of our DSL and we explain the main components of the SOS specification, involving types of configurations and evaluation rules both for expressions over base types and for DSL statements. The DSL only includes the set of model actions `create`, `set` and `unset`, the no-op construct (), `let`-binding and the operation `snapshot` $x$ $\{s2\}$, which applies the sequence $s2$ of model actions to the set of properties of the object referenced by $x$. In our language, the `snapshot` operator is used to focus the interpreter on an object and to manipulate it locally. This command is useful when applying sizeable bulks of model actions to an object in a large model without having to traverse the model for each action.

*Definition 9.* (DSL Syntax) Given a model $\mathcal{M} = (\mathcal{B}, c, \mathcal{P}, cl, df, oe, sr, r)$ and a set **Var** of variables names, the syntax of our DSL is given by sentences of the syntactic category **Stmt** as follows:

$$\begin{aligned}
\mathbf{Stmt} \ni s ::=\ & \mathtt{snapshot}\ x\ \{s2\}\,|\,\mathtt{let}\ x = e\ \mathtt{in}\ s \\
& |\ \mathtt{let}\ x = c\ \mathtt{in}\ s\,|\,\mathtt{create}(c) \\
& |\ \mathtt{delete}(x)\,|\,s; s\,|\,() \qquad \text{(statements 1)} \\
\mathbf{ActStmt} \ni s2 ::=\ & \mathtt{let}\ x = e\ \mathtt{in}\ s2 \\
& |\ \mathtt{let}\ x = \mathtt{create}(c)\ \mathtt{in}\ s2 \\
& |\ \mathtt{create}(p,c)\,|\,\mathtt{set}(p,x)\,|\,\mathtt{setCmt}(p,x) \\
& |\ \mathtt{unset}(p)\,|\,\mathtt{unset}(p,x) \\
& |\ s2; s2\,|\,\bullet \qquad \text{(statements 2)} \\
\mathbf{Expr} \ni e ::=\ & v\,|\,x \qquad \text{(base type expressions)}
\end{aligned}$$

where $x \in \mathbf{Var}$ denotes a variable that can be bound to a value $v$ of a base type $b$ in $\mathcal{B}$, and **Expr** corresponds to expressions built over base data types.

As a running example we use the implementation of the effects of the pull up field refactoring rule, adapted from [34].

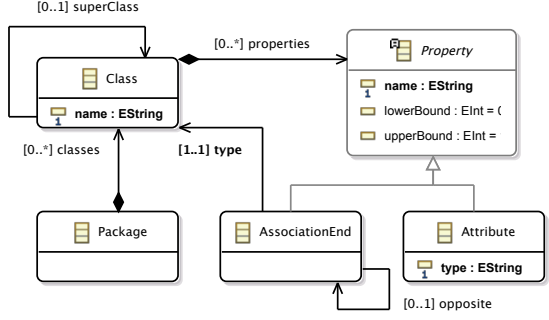

Figure 2: Simple metamodel for class diagrams.

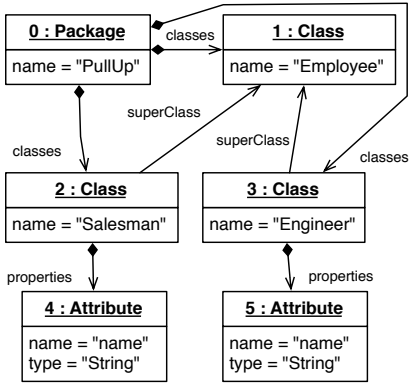

Figure 3: Class diagram.

We consider refactorings in class diagrams according to the language defined by the metamodel in Fig. 2.

A program in our DSL to implement the effects of the pull up field refactoring rule on the example of Fig. 3 could be composed as follows:

```
let var("0") = oid("1") in
let var("1") = oid("3") in
let var("2") = oid("4") in
let var("3") = oid("5") in
snapshot var("0") {
  setCmt("properties", var("1"))
};
snapshot var("2") {
  unset("properties", var("3"))
}
```

where the let statements declare the object identifier variables to be used in the statements; the first `snapshot` command pulls up the attribute `name` by *moving it*, without creating copies, from object identified by `oid("1")` to object `oid("0")`; and the last `snapshot` command gets rid of the object representing the attribute `name` in object `oid("2")`. As we discuss in 3.5, this program preserves the correctness of the resulting class diagram.

## 3.3 FMA Semantics

The SOS of our DSL is specified in terms of two types of configurations, representing the interpreter state for executing first-level and second-level FMA statements, and of one evaluation relation for each of them, specifying the semantics of each statement. In this subsection, we present the types of configurations that are used in the FMA's SOS and their components. Configurations are then used to define the evaluation relation. The formal definitions of notions and operations involved can be found in [10, Appendix B].

### 3.3.1 Configurations and Environments

In a configuration for our interpreter, an environment $\eta$ is defined as the disjoint union of: partial functions between variables and values for each base type in $\mathcal{B}$; a partial function between object identifiers and their locations; and a function $new : \mathcal{C} \to \mathcal{O}$ mapping each class name to a fresh identifier. We can also query information from the environment by using the expressions $\eta(x)$ to obtain the value of variable $x$, $\eta(o)$ to obtain the location of object identifier $o$, $\eta(c)$ to obtain the object identifier of class name $c$.

We represent specific configurations by tuples of the form $\eta \,|\, os \,|\, s$ for first-level statements and $\eta \,|\, os \,|\, l \,|\, ps \,|\, as \,|\, s2$ for second-level statements, where: $\eta$ is the environment of variables, references and fresh identifiers; $os$ is the structured model; $l$ points to the location in the model that currently receives the focus of the interpreter; $ps$ is the set of properties of the object under the focus of the interpreter; $as$ is the ordered set of deferred actions; and $s$ and $s2$ is the statements to be evaluated.

Given an input model $M = (os, \Pi)$, an initial configuration is defined as follows: the environment $\eta$ is initialized with the locations in $os$ for each object identifier and the map $new$ is also initialized for each class name that appears in the model; the object set $os$ is taken as the model; $l$ is set to $\rho_o$; the set of properties $ps$ is left empty; and finally the statement $s$ corresponds to the program to be evaluated. The terminal configurations for first-level statements are defined as tuples of the form $(\eta \,|\, os \,|\, s)$, where $s = ()$, and for second-level statements as $(\eta \,|\, \rho \,|\, l \,|\, ps \,|\, s2)$, where $s2 = \bullet$.

As we are dealing with structured models, we can uniquely identify the location of an object $\rho'$ as a path from the root object $\rho$ of the model to $\rho'$ by traversing containment properties, i.e. properties bound to sets of objects. An object location is a path $l$ formed by a sequence of pairs formed by object identifiers and containment reference names from the root object to the object under focus, which is suffixed by the identifier of the object under focus.

To specify the semantics of FMA we need two auxiliary operations to query and update structured models at specific locations. Given an object $\rho$, for $l \in \mathbf{Loc}_\rho$:

- $os|_l$ produces the pair $\langle os' \,;\, \rho \rangle$ formed by the object $\rho$ at location $l$ inside the structured object $os$ and by the complex object $os'$, which equates to the object set $os$ after extracting the object $\rho$.

- $os[\rho]_l$ inserts an object $\rho$ at a new location in a host object set $os$ forming a new object set $os'$.

### 3.3.2 Evaluation Relation

In the evaluation relations, transition are of the form

$$\eta \,|\, os \,|\, s \Downarrow_s \eta' \,|\, os' \,|\, s'$$

for first-level statements and of the form

$$\eta \,|\, os \,|\, l \,|\, ps \,|\, as \,|\, s2 \Downarrow_{s2} \eta' \,|\, os' \,|\, l' \,|\, ps' \,|\, as' \,|\, s2'$$

for second-level statements. The complete specification is given in [10, Appendix B.3]. In the following, we illustrate excerpts of these two evaluation relations.

The evaluation relation for second-level statements specifies how to apply model actions in a specific object by acting on its set of properties. When the property being updated is a bidirectional reference, model actions generate a finite number of additional model actions with the purpose of updating the opposite reference. Deferred model actions are executed in the `snapshot` operator. In the following we present the semantic rules for the model actions that deal with references (non-containment), shown in Figure 3:

- The model action $set(p, x)$ adds the reference $x$ to property $p$ in the set of properties of the object under focus. Rule E-REFUNISET applies when a reference is unidirectional, i.e. there is no opposite end in $\mathcal{M}_{oe}$ and rule E-REFBISET applies when the reference is bidirectional. Both rules are defined when the object to be referenced exists in the model, that is, if its object identifier is mapped to a location in the environment $\eta$. Both rules insert the object identifier at the end of the ordered set if the collection $is$ does not contain it, preserving the set semantics of the collection.

- The model action $unset(p, x)$ deletes the reference $x$ from the set of identifiers contained by the property $p$ in the set of properties of the object under focus. Its semantics is specified by the rule E-REFUNIUNSET for unidirectional references and by the rule E-REFBIUNSET for bidirectional references. In the second case, a new action is created in order to unset the opposite reference. Its application is deferred until the end of the execution of the containing snapshot statement.

The evaluation relation for first-level statements specify the semantics of FMA statements. We illustrate this relation by focussing on the semantic rule for the `snapshot` command, shown in Figure 4.

The operator `snapshot x {s2}` performs side effects on the model instance by replacing the property set, in the object identified by the reference $x$, with the property set that results from the application of the sequence of model actions $s2$. The execution specification of a statement in the `snapshot x {s2}` in the rule E-SNAPSHOT changes the focus of the interpreter to the object referenced by $x$ so that the intermediate execution steps contained in block $s2$ continue with this focus. This is achieved by updating the location $l$ and the property set $ps$ in the configuration of the intermediate evaluation transitions. This avoids traversals in the structured model every time it needs to be updated and all the effects are combined in the property set. These effects are realized on the model when the interpreter finishes the evaluation of the containing `snapshot` statement as explained above. In addition, intermediate transitions proceed with the model instance $os'$ resulting from extracting the object under focus $\rho''$ in the side condition.

At the end of the transition, the new model instance $os'''$ is updated by inserting at location $l'$ the object $\rho''$ with the effects resulting from model actions in intermediate transitions. The expression $eval(as, \eta', os'''[<\rho_o''|ps'>]_{l'})$ applies the deferred model actions $as$ in a model in order to keep the consistency of bidirectional references. The focus is restored once the evaluation transition is completed.

## 3.4 Validity of Model Transformations

In this section, we show that the SOS specification of FMA characterizes programs that are safe with respect to the notion of valid model, i.e. the validity of model transformations. In other words, given a valid model, the execution of a FMA program will always result in a valid model, or else it will return a trapped error. In addition, the semantics of FMA is deterministic and terminating as shown in [10, Appendix C].

In what follows, we provide the scaffolding required to prove that FMA is a safe language with respect to valid models, which amounts to show its totality and the subject reduction property of the evaluation relation. That is, that execution cannot get stuck due to a trapped error and that all models that can be computed are valid. To simplify the formulation of results and their proofs, we are going to develop them under the assumption that trapped errors do not occur[8].

A closed FMA program is a first-level statement where all contained variables are bound, i.e. where there are no free variables.

*Definition 10.* Configuration with $M$ and $s$. A configuration with a model $M = (os, \Pi)$ of model type $\mathcal{M}$ and a closed FMA program $s$ is a configuration $k = \eta \mid os \mid s$ where

$$\eta = VM \uplus loc(os) \uplus fresh(M)$$

where where $VM \subset (\mathbf{Var} \rightharpoonup \mathcal{D}_b \uplus \mathbf{Var} \rightharpoonup \mathcal{O})$, $loc(os)$ obtains a map from the identifier of each object in $M$ to its location, and $fresh(M)$ obtains a map from class names in $\mathcal{M}_C$ to a fresh identifier.

*Theorem 1.* Totality of the evaluation relation $\Downarrow_s$. Given a configuration $k$ with model $M$ and first-level statement $s$, where $M : \mathcal{M}_{pre}$ and a configuration $k'$, if $k \Downarrow_s k'$ then $k$ is in normal form, in which case $k_s' = ()$.

PROOF. (Idea) By induction on the structure of the FMA programs, show that for each first-level configuration with model $M$ and a statement $s$ that is different from () we can apply a SOS rule that leads to the desired result. The proof requires a lemma stating a similar result for second-level statements. □

*Theorem 2.* $\Downarrow_s$ preserves model validity. Given a configuration $k$ with model $M$ and a closed FMA program $s$, and a configuration $k'$ with model $M'$, if $M : \mathcal{M}_{pre}$ and $k \Downarrow_s k'$ then $M' : \mathcal{M}_{pre}$.

To prove this theorem, we introduce a lemma to show that $eval$ returns a valid model after applying model actions to an object in the `snapshot` command.

*Lemma 1.* Given $M = (os, \Pi)$ such that $M : \mathcal{M}_{pre}$, a configuration $k$ with $M$, where $k = \eta \mid os \mid s$ for a closed FMA program $s$, and a second level statement $s2$ closed with respect to $\eta$, if $< os'; \rho > = os|_l$ and

$$\eta \mid os' \mid l \mid \rho_{ps} \mid \emptyset \mid s2 \Downarrow_{s2} \eta' \mid os' \mid l'' \mid ps' \mid as \mid \bullet$$

then $eval(as, \eta', <\rho_o \mid ps'>[os']_l) : \mathcal{M}_{pre}$.

---
[8]The formal development of results, including a treatment of trapped errors, are considered in [10, Appendix D].

$$\frac{noOp(bRE(last(l), p)) \quad \eta \vdash x \Downarrow_e o \quad l' = \eta(o)}{\eta \,|\, os \,|\, l \,|\, p = is, ps \,|\, as \,|\, \texttt{set}(p, x) \Downarrow_{s2} \eta \,|\, os \,|\, l \,|\, p = (o \in is)? \, is : is \, o, ps \,|\, as \,|\, \bullet} \quad \text{(E-RefUniSet)}$$

$$\frac{bRE(c', p') = getOp(bRE(last(l), p)) \\ \eta \vdash x \Downarrow_e o \quad l' = \eta(o) \quad is' = (o \in is)? \, is : is \, o \quad a = set(o, p', last(l))}{\eta \,|\, os \,|\, l \,|\, p = is, ps \,|\, as \,|\, \texttt{set}(p, x) \Downarrow_{s2} \eta \,|\, os \,|\, l \,|\, p = is' \,|\, (o \in is)? \, as : as \, a \,|\, \bullet} \quad \text{(E-RefBiSet)}$$

$$\frac{noOp(bRE(last(l), p)) \quad \eta \vdash x \Downarrow_e o}{\eta \,|\, os \,|\, l \,|\, p = o \, is, ps \,|\, as \,|\, \texttt{unset}(p, x) \Downarrow_{s2} \eta \,|\, os \,|\, l \,|\, p = is, ps \,|\, as \,|\, \bullet} \quad \text{(E-RefUniUnset)}$$

$$\frac{bRE(c', p') = getOp(bRE(last(l), p)) \quad \eta \vdash x \Downarrow_e o \quad a = unset(o, p', last(l))}{\eta \,|\, \rho \,|\, l \,|\, p = o \, is, ps \,|\, as \,|\, \texttt{unset}(p, x) \Downarrow_{s2} \eta \,|\, \rho \,|\, l \,|\, p = is, ps \,|\, as \, a \,|\, \bullet} \quad \text{(E-RefBiUnset)}$$

Table 3: SOS of reference actions.

$$\frac{\eta \vdash x \Downarrow_e o \quad l' = \eta(o) \quad \langle os', \rho'' \rangle = os|_{l'} \quad \eta \,|\, os' \,|\, l' \,|\, \rho''_{ps} \,|\, \emptyset \,|\, s2 \Downarrow_{s2} \eta' \,|\, os''' \,|\, l'' \,|\, ps' \,|\, as \,|\, \bullet}{\eta \,|\, os \,|\, \texttt{snapshot} \, x \, \{s2\} \Downarrow_s \eta' \,|\, eval(as, \eta', os'''[{<}\rho''_o \,|\, ps'{>}]_{l'}) \,|\, ()} \quad \text{(E-FmaSnapshot)}$$

Table 4: SOS of first-level statements.

PROOF. (Idea) The proof consists in showing that the two defining conditions of the notion of valid (structured) model, namely referential integrity and consistency of bidirectional references, are preserved by the application of second-level statements, including model actions. The cases of relevance in these proofs are summarized as follows:

- The avoidance of parallel edges is ensured by the post-conditions of the rules E-RefUniSet, E-RefBiSet, E-CmtUniSet, E-CmtBiSet, by specifying that duplicate references cannot be added to an ordered set of references and that duplicate objects cannot be added to a set of contained objects.

- the creation of dangling edges is avoided by forbidding the application of the rules E-RefBiSet, E-RefBiSet, E-CmtUniSet and E-CmtBiSet when the referenced object does not have a concrete location in the model.

- the creation of inconsistent opposite references is avoided by updating opposite references in the rules E-RefBiSet, E-CmtBiSet, E-RefBiUnset and E-CmtUnset, which are executed by the rule E-Snapshot, when one of the two references is set or unset.

□

PROOF. Theorem 1 (Idea). The proof follows by induction on the structure of FMA programs, using Lemma 1 for the rule E-Snapshot. □

### 3.5 Well-Behaved Model Transformations

In this section, we show that well-typed, closed FMA programs correspond to well-behaved model transformations. Given a consistent model, the model transformation dictated by a well-typed, closed FMA program will always produce a consistent model in the absence of trapped errors. We present a type system for FMA programs and show that it is sound with respect to the notion of model consistency.

Since the semantics of FMA programs is deterministic, terminating, and totally defined, the consistency theorem corresponds to total correctness of closed FMA programs with respect to consistent models in general, and not just for a specific model transformation program. More informally, this means that metamodel-conformant model transformations *never go wrong*, and run-time errors that cause

$$\frac{\Gamma \,|\, \Pi \vdash x : ref \, c \quad c \,|\, \Gamma \,|\, \Pi \vdash s2 : ()}{\Gamma \,|\, \Pi \vdash \texttt{snapshot} \, x \, \{s2\} : ()} \quad \text{(T-FmaSnapshot)}$$

Table 5: Typing rule for the snapshot command.

$$\frac{\Gamma \,|\, \Pi \vdash e : \alpha \quad c \,|\, \Gamma \,|\, \Pi \vdash (p : \alpha) <: \mathcal{M}_{cl}(c)|_p}{c \,|\, \Gamma \,|\, \Pi \vdash \texttt{set}(p, e) : \bullet} \quad \text{(T-Set)}$$

Table 6: Typing rules for second-level set statements.

inconsistencies in a model do not occur. Therefore, FMA's type checker can be seen as a pragmatic tool for developing transformations since the tool provides feedback to the user while writing the transformation.

The typing relations presented in Table 1 and in Table 2 for model types are augmented for FMA statements. Given a typing environment $\Gamma$ for variables and a typing environment $\Pi$ for references, three new kinds of typing judgements are introduced: $\Gamma \,|\, \Pi \vdash e : \alpha$ for expressions; $\Gamma \,|\, \Pi \vdash s : \tau$ for first-level statements; $c \,|\, \Gamma \,|\, \Pi \vdash s : \tau$ for second-level statements, where $c$ corresponds to the class name of the object under focus.

The typing rule for the snapshot first-level statement, extracted from the typing relation for first-level statements and shown in Table 5, specifies that a snapshot command is well-typed if the expression $x$ corresponds to an object reference and if the block $s2$ of second-level statements is well typed. The typing rule for the set model action that modifies an attribute, a reference or a containment, extracted from the typing relation for second-level statements, is shown in Table 6. The rule (T-Set) specifies that a statement $set(p, e)$ is well-typed when the type of the given expression $e$ is compatible with the type of the property $p$ in class $c$, i.e. the property type is direct type of the expression type or a supertype. Technically, the property field with name $p$ is projected from the type $\mathcal{M}_{cl}(c)$ by using the domain restriction $\mathcal{M}_{cl}(c)|_p$.

The following theorem states that evaluation relation for FMA programs preserves the type of the models being transformed.

*Theorem 3.* Consistency for $\Downarrow_s$. Given a structured model $M = (os, \Pi)$, and a configuration $k$ with $M$ and a well-typed, closed FMA statement $s$, if $M : \mathcal{M}$ and $k \Downarrow_s k'$ then, for some $\Pi' \supseteq \Pi$, $M' = (k'_{os}, \Pi')$ is a structured model and $M' : \mathcal{M}$.

PROOF. (Idea) By induction on the typing relation, analyzing the cases for the last step of a typing derivation. Showing the soundness of the typing rule for the snapshot statement requires proving that replacing a modified object in the model is consistent, which can be done following the similar strategy followed in the proof of the validity theorem for first-level statements. □

Given a consistent model $M : \mathcal{M}$, an FMA program $s$ is a metamodel-conformant model transformation that represents a well-behaved model transformation for that model iff $\emptyset \mid M_\Pi \vdash s : ()$. This follows from Theorem 1 and from Theorem 3. Therefore, the type system specifies a decision procedure for determining whether a FMA program is well behaved with respect to the notion of structured model consistency.

## 4. RELATED WORK

The increasing role of model transformations as model compilers has lead to a reasonably high activity of research on the verification of model transformations in the MDE community, as evidenced by a number of recently published surveys [2, 4, 29]. We have already provided a discussion on related work about model subtyping in Subsection 2.1. In what follows we revise a number of approaches that focus on the verification of model transformations from a program correctness perspective, focussing on type checking approaches. A discussion on related approaches for the semantics of deterministic model transformations can be found in [10].

### 4.1 Correctness of Model Transformations

Preservation of transformation intent [44], defined as a set of properties, is identified as a key factor common to most reuse approaches. In [49], the authors address the validity, with respect to transformation intent, of a type compatibility relation that determines when a model type is substitutable for another one, both for subtyping, using automated mechanisms, and for concept-based adaptation, using manual intervention. In our approach, the transformation intent is captured by a sound type that ensures that well-typed model transformations are well behaved with respect to the notion of consistent (structured) model. The implementation of our type system provides a lightweight mechanism for ensuring the correctness of model transformations without requiring further proofs, on the side of the programmer.

Correctness of graph transformations ensured by generation of weakest preconditions was first proposed in [39], where the idea is to generate application conditions that lead to a correct application of a graph transformation rule with respect to a postcondition graph constraint. This technique has been extended to symbolic graphs in [28] and to a representative excerpt of OCL constraints in [21]. In the latter, Clarisó et al. show how to compute weakest preconditions for graph transformation rules using backwards reasoning that can be used to verify correctness of model transformations against very informative model types, that is, enriched with OCL constraints. The underlying motivation and philosophy of this method is quite different, and complementary in practice, to ours: their work focusses on finding a model that satisfies the preconditions of a sequence of rules in order either to produce a valid result w.r.t. given postconditions when the sequence is fixed (white-box testing scenario) or not (backwards reachability), or to produce an unwanted result (deadlock analysis)[9]; our work can be used to *check* that the sequence of actions (corresponding to graph transformation rules) are indeed correct without the need of performing simulations, which can be used for gaining confidence on the conclusions of their inferences in all the scenarios above.

Hoare-style partial correctness of ATL model transformations, where correctness is considered with respect to OCL constraints of the involved metamodels, is addressed in [16] using Alloy bounded model verification tool via UML2Alloy and in [15] using SMT solvers. Their methodology consists in reducing the problem of verifying rule-based transformations between constrained metamodels to the problem constrained metamodels only. So a transformation model is expressed as a metamodel that integrates the source and target metamodel of the transformation together with constraints that specify the transformation intent. The verification consists in negating a constraint of the target metamodel in order to search for counterexamples. If none is found, that property is satisfied. In [16], transformation models have to be bound to ensure decidability at the expense of not finding bugs, which may sit out of the scope of the search space. In [15], ATL transformations are translated into first-order logic and symbolic reasoning is used to check the satisfaction of the transformation intent expressed in constraints. This approach does not require bounds on the model extent although it is incomplete (in the sense that there are properties that cannot automatically decided). So the latter can be used to verify several pre-post implications, but is not suitable to find counter examples.

Safety of model transformations is studied from a refinement perspective in [17] and [58], where model transformations are the artifacts under refinement. We are, however, interested in enabling the reuse of model transformations for refined model types.

Syntactic correctness of model transformations is addressed in [59] for model transformations (at the model level) as a model finding problem and for model transformation systems (at the metamodel level) as a theorem proving problem. The authors provide a mechanism to prove correctness of model transformations by implementing a proof system through planner algorithms. Models are encoded as prolog-style facts and transformations as planner operations. The proof of correctness for a particular model transformation system is given indirectly by delegating it to the effectiveness of the planner algorithms used. This proof method is powerful given that more sophisticated theorem provers can be substituted for the planning algorithms (which could also be used to prove other properties).

### 4.2 Static Type Checking

Static type checking of model transformation programs in VIATRA2 was studied by Ujhelyi et al. in [57, 56] where the type checking problem is reduced to a constraint satisfaction problem. Given a model transformation in Viatra2 (either in VTCL or in GT), from the abstract syntax tree of a program, the tool generates a type constraint for each expression and statement in the program, and the resulting constraint satisfaction problem is used to find bug patterns (analysis problems, inconsistencies, traversal problems). In this work, the authors analyze the performance of the resulting type checker but the semantic soundness of the type system is not addressed. This means that the tool can detect common bug patterns (known at design time of the type system) following a pragmatic approach: the type system can be refined with more bug patterns as practice with the model transformation language leads to more expertise and knowledge so that more expressive model transformations can be statically analysed. Hence, although the type checker is useful for early detection of run-time errors and helps to improve the quality of model transformations, it still falls under the category of weakly-typed languages, since it cannot guarantee that well typed programs are well behaved. That is, it guarantees the absence of bugs detected by patterns but not the absence of bugs w.r.t. typing constraints in general.

---

[9]The authors also show an additional scenario for checking rule independence.

In [24], Cuadrado et al. present a static type checking approach for ATL programs. Their method is implemented in three main phases: the ATL transformation is statically typed checked using an OCL type inference system that identifies potential bugs; a transformation dependence graph is extracted and analysed in order to find potential bugs regarding the application of rules; and a final phase resolves their authenticity by using model finders in order to discover a witness of the bug or else to discard the error. The type system used in [24] considers OCL expressions but its semantic soundness has not been shown yet.

Zschaler also considers types as constraints in [61] under the notion of constrained model type. They provide a type system for the static analysis of model management programs using a white-box approach, that is statements that yield fine-grained side-effects on a model can be analysed. Their work does not discuss whether their type system ensures the safe application of model management operations or of their reuse in the presence of model subtypes.

George et al. discuss the use of Scala for developing a DSL for implementing model transformations [36]. The authors show that Scala's type inference mechanism facilitates type checking of model transformation programs, similar to ATL or RubyTL. However, the type system has not been shown to be semantically sound.

## 5. CONCLUSIONS

In the first part of the paper, we have discussed current approaches to model reuse and the lack of approaches, to the best of our knowledge, that deal with model subtyping from a subsumption perspective. We have proposed a type system for models based on structural type inference that provides support for polymorphic subtyping (subsumption). The type system formalizes the notions of model type, of metamodel conformance and of model subtyping and provides support for implicit, duck typing for model transformation and model management languages.

In the second part of the paper, the notion of EMF model graph has been decomposed into the notions of valid (structured) model and of consistent (structured) model to facilitate the study of safety in model transformations and of semantic soundness in type systems for model transformation languages. The most common effects on models resulting from model transformations have been formalized in a DSL of focussed model actions, by providing a big-step structural operational semantics. The DSL has been shown to be safe with respect to valid (structured) models. The type system for models has been augmented in order to support type checking of FMA programs, characterizing well-behaved model transformations with respect to the notion of consistent (structured) models. Therefore, FMA is a strongly-typed model transformation language and metamodel-conformant model transformations never go wrong. Apart from model transformation languages based on graph transformation theory, most current model transformation languages are weakly typed, as discussed in subsections 2.1 and 4.2.

The semantics of FMA has been implemented as an interpreter for the model transformation language. The type system for models and for FMA programs has also been implemented, providing an executable decision procedure to check when a model transformation is metamodel conformant. The prototypes are available online[10]. A variant of the interpreter has been applied to implement the model transformation language JSON upsert-delete[11], used to execute safe model transformations over models represented in JSON format.

In this paper, we have focussed on the action side of model transformations, by providing a semantics for safe model actions. However, a model transformation language usually provides more expressive constructs, such as pattern matching, queries, control structures and procedural abstractions. The augmentation of FMA with those language constructs is a natural extension of this work given that the semantics of the language is compositional.

Regarding reuse of model transformations, the notion of model subtyping based on subsumption suffers from being too liberal in contexts where all the substitutions need to be analyzed. The inclusion of a more prescriptive approach that limits the amount of valid substitutions or that dictates how substitutions should be performed is another potential extension of the current type system.

## Acknowledgements

The author thanks the anonymous referees of SLE'15 and SLE'16 for their helpful comments on a previous draft of this document.

---

[10] http://arturboronat.info/fma
[11] http://arturboronat.info/jsonud

# APPENDIX

## A. TYPING RELATION FOR MODELS

### A.1 Meta-properties of the Typing Relation

*Lemma 2.* (Unique typing) Given typing assumptions $\Gamma$ for variables and $\Pi$ for object identifiers, for all terms $t$ there is at most one type $\tau$ such that $\Gamma; \Pi \vdash t : \tau$.

PROOF. The sketch of the proof is as follows. This lemma can be proved in two steps: first we prove that each rule determines a unique type for a given term by using induction on the structure of the terms involved in the typing judgements; second, for those rules that overlap in their premises we analyse whether all critical pairs can be reconciled.

The base cases correspond to the axioms of the type system for base values, undefined references, undefined containments, and empty property sets, and by direct proof the type inferred can be shown to be unique. The induction step cases correspond to the other rules as follows:

- rules (T-REF1) and (T-REF2): these two rules need to be analysed together since their premises overlap and, for the same term, they may diverge when we have that $C, C' \in \mathcal{C}$ and $cl(C) <: cl(C')$ is satisfied for (T-REF1) and $cl(C') <: cl(C)$ is satified for (T-REF2). Since $cl$ is injective, these two cases can only occur when $C = C'$. Hence, the inferred type is unique.

- rules (T-OBJ1) and (T-OBJ2): the proof follows an argument similar to the one used for the rules (T-REF1) and (T-REF2).

- rule (T-PROP): by composing the types from the premises, assumed to be unique, we get a new type.

- rule (T-OBJ): the injectivity of $cl$ guarantees the uniqueness of the inferred type by direct proof.

□

### A.2 Decidability of the Subtyping Relation

The object subtyping relation has been described using recursion in Definition 4 in order to abstract technical detail in its presentation. In order to use the typing relation a decision procedure for type checking, the function defining the relation is equipped with a small memory that flags when two class names have been compared once. Once this is the case, the function assumes that the relation holds for the two class names.

The complete definition of the function defining the relation is as follows:

$subtypeOf(\varsigma_1, \varsigma_2, \text{SCS})$
$= \textit{true if } \varsigma_2 = \emptyset$
$= \textit{false if } \varsigma_1 = \emptyset \land not(\varsigma_2 = \emptyset)$
$= \textit{true if } \varsigma_1 = \bot$
$= \textit{false if } not(\varsigma_1 = \emptyset) \land (\varsigma_2 = \bot)$
$= (b_1 = b_2)? \; subtypeOf(\varsigma_1', \varsigma_2', \text{SCS}) : \textit{false}$
$\quad \textit{if } \varsigma_1 = (p : b_1), \varsigma_1' \land \varsigma_2 = (p : b_2), \varsigma_2'$
$= ((c_1, c_2) \in \text{SCS})?$
$\quad subtypeOf(\varsigma_1', \varsigma_2', \text{SCS}) :$
$\quad (subtypeOf(cl(c_1), cl(c_2), \{(c_1, c_2)\} \cup \text{SCS}))?$
$\quad\quad subtypeOf(\varsigma_1', \varsigma_2', \text{SCS}) :$
$\quad\quad \textit{false}$
$\quad \textit{if } \varsigma_1 = (p : ref \; c_1), \varsigma_1' \land \varsigma_2 = (p : ref \; c_2), \varsigma_2'$
$= subtypeOf(\varsigma_1', \varsigma_2', \text{SCS})$
$\quad \textit{if } \varsigma_1 = (p : ref \; \texttt{Any}), \varsigma_1' \land \varsigma_2 = (p : ref \; c_2), \varsigma_2'$
$= ((c_1, c_2) \in \text{SCS})?$
$\quad subtypeOf(\varsigma_1', \varsigma_2', \text{SCS}) :$
$\quad (subtypeOf(cl(c_1), cl(c_2), \{(c_1, c_2)\} \cup \text{SCS}))?$
$\quad\quad subtypeOf(\varsigma_1', \varsigma_2', \text{SCS}) :$
$\quad\quad \textit{false}$
$\quad \textit{if } \varsigma_1 = (p : c_1), \varsigma_1' \land \varsigma_2 = (p : c_2), \varsigma_2'$
$= subtypeOf(\varsigma_1', \varsigma_2', \text{SCS})$
$\quad \textit{if } \varsigma_1 = (p : \texttt{Any}), \varsigma_1' \land \varsigma_2 = (p : c_2), \varsigma_2'$
$= subtypeOf(\varsigma_1', \varsigma_2, \text{SCS}) \textit{ where } \varsigma_1 = (p : \alpha), \varsigma_1' \textit{ otherwise}$

The object subtyping relation from Definition 4 is refined in terms of function $subtypeOf$ as follows:

*Definition 11.* Object Subtyping (Subsumption)
$$\varsigma_1 <: \varsigma_2 \iff subtypeOf(\varsigma_1, \varsigma_2, \emptyset)$$

## B. FMA SOS SPECIFICATION

### B.1 Object Locations and Environments

To be able to specify the semantics of our DSL, we need to introduce two important notions: object locations and environments.

As we are dealing with structured models, we can uniquely identify the location of an object $\rho'$ as a path from the root object $\rho$ of the model to $\rho'$ by traversing containment properties, i.e. properties bound to sets of objects. An object location is a path $l$ formed by a sequence of pairs formed by object identifiers and containment reference names from the root object to the object under focus, which is suffixed by the identifier of the object under focus.

*Definition 12.* (Object Locations) The set **Loc** of object locations $l$ is defined as

$\mathcal{O} \cup \{o.p.l \mid o \in \mathcal{O}, l \in \textbf{Loc},$
$\quad \exists \rho \in \textbf{Object}(\rho_o = o, \exists (p = \rho'' \; os) \in \rho_{ps}(\rho_o'' = l_h))\}.$

*Definition 13.* (Object Locations in an Object) The set $\textbf{Loc}_\rho$ of locations in an object $\rho$ is defined as follows:

$$\{l \in \textbf{Loc} \mid l_h = \rho_o\}.$$

To specify the semantics of the `snapshot` command we need two auxiliary operations to query and update structured models at specific locations. Given an object $\rho$, for $l \in \mathbf{Loc}_\rho$, the pair $\langle \rho' \,;\, \rho'' \rangle$, denoted by $\rho|_l$, formed by the object $\rho''$ at location $l$ inside the complex object $\rho$ and by the complex object $\rho'$, which equates to the object $\rho$ after extracting the object $\rho''$, is defined by induction on the length of $l$ as follows:

$$(\texttt{<o|ps>}, os)|_o = \langle os \,;\, \texttt{<o|ps>} \rangle$$
$$(\texttt{<o'|ps'>}, os)|_o =$$
$$\qquad \texttt{<o'|ps'>} os|_o \quad \text{if } o \neq o'$$
$$(\texttt{<o|p = os, ps>}, os')|_{o.p.l} =$$
$$\qquad \texttt{<o|p = } os|_l\texttt{, ps>}, os'$$
$$\texttt{<o|p = } \langle os \,;\, \rho \rangle \, os'\texttt{, ps>} =$$
$$\qquad \texttt{<o|p = } \langle os\, os' \,;\, \rho \rangle\texttt{, ps>} \quad \text{if } os' \neq \emptyset$$
$$\texttt{<o|p = } \langle os \,;\, \rho \rangle\texttt{, ps>} =$$
$$\qquad \langle \texttt{<o|p = os, ps>} \,;\, \rho \rangle$$

A second operation that we will need in our semantics inserts an object $\rho$ at a new location in a host object $\rho'$ forming a new complex object $\rho''$. The set $\mathbf{Loc}_{\rho[\rho']}$ of new locations for the object $\rho'$ in the object $\rho$ is defined as

$$\{l \in \mathbf{Loc}_\rho \cup \mathcal{O} \mid \text{either } l = \rho'_o \text{ or, for } l' \in \mathbf{Loc}_\rho, \rho'' \in \mathbf{Object},$$
$$\text{we have both that } (p = os) \in \rho''_{ps}$$
$$\text{and that } l = l'.p.\rho'_o\}$$

For $l \in \mathbf{Loc}_{\rho[\rho']}$, we denote by $\rho[\rho']_l$, the object resulting from inserting the object $\rho'$ in object $\rho$ at location $l$ as follows:

$$(\texttt{<o|ps>}, os)[\rho]_o = \texttt{<o|ps>}, os$$
$$os[\rho]_o = \rho, os \quad \text{where } \rho_o = o$$
$$(\texttt{<o|p = os, ps>}, os')[\rho]_{o.p.l} =$$
$$\qquad \texttt{<o|p = } os[\rho]_l\texttt{, ps>}, os'$$

Note that for this operation to be well defined, the identifier of the object to be inserted must coincide with the last element of the location at which the object is to be placed. If the location at which the object $\rho$ is to be inserted is already taken, the insertion does not take place. Moreover, the two operations above are defined for sets of objects, which may be singleton sets.

In a configuration for our interpreter, an environment $\eta$ is defined as the disjoint union of: partial functions between variables and values for each base type in $\mathcal{B}$; a partial function between object identifiers and their locations; and a function $new : \mathcal{C} \to \mathcal{O}$ mapping each class name to a fresh identifier.

*Definition 14.* (Environments) Given a model $\mathcal{M} = (\mathcal{B}, \mathcal{C}, \mathcal{P}, <:, M)$, the set $\mathbf{Env}$ of environments $\eta$ is defined as

$$\bigsqcup_{b \in \mathcal{B}} (\mathbf{Var} \rightharpoonup \mathcal{D}_b) \sqcup \mathbf{Var} \rightharpoonup \mathcal{O} \sqcup \mathcal{O} \rightharpoonup \mathbf{Loc} \sqcup \mathcal{C} \to \mathcal{O}$$

Note that for variables of base types we are abstracting away the store so that we do not have to deal with aliasing problems. To work with environments we use the notation $\eta[x \mapsto v]$, $\eta[o \mapsto l]$, $\eta[c \mapsto o]$, to denote an update of a variable, identifier or class name mappings, respectively, in the environment $\eta$. We can also apply several updates, e.g. two updates $x \mapsto v$ and $x' \mapsto v'$, using the notation $\eta[x \mapsto v; x' \mapsto v']$. Furthermore, whenever the corresponding variable, identifier or class name does not exist in the environment, the update represents the insertion of the corresponding mapping in the environment. On the other hand, we can also query information from the environment by using the expressions $\eta(x)$ to obtain the value of variable $x$, $\eta(o)$ to obtain the location of object identifier $o$, $\eta(c)$ to obtain the object identifier of class name $c$. Given an environment $\eta$, each of its components can be projected by using the unary operators $\_|_{\mathbf{Var}}$, $\_|_{\mathbf{OidVar}}$, $\_|_{\mathbf{Loc}}$, $\_|_{\mathbf{New}}$.

## B.2 Expressions

In our DSL, the definition of the set $\mathcal{B}$ of base type names is dependent on each target implementation platform so that expressions over base types can be formed by using the operations provided by the corresponding built-in base types.

The set $\mathcal{K}_e$ of configurations for evaluating expressions is defined as $(\mathbf{EnvP} \times \mathbf{Env} \times \mathbf{Expr}) \cup \bigcup_{b \in \mathcal{B}} \mathcal{D}_b$.

An evaluation transition in our semantics for expressions is of the form $\eta \vdash e \Downarrow_e v$, where $\eta$ is the environment of declarations of variables, object identifiers and fresh identifiers for class names, and $v$ is the value bound to a variable. Note, however, that we are dealing with value representations and that their interpretation is left implicit.

The structural operational semantics for expressions is given by the rules in Table 7, where the axiom E-VALUE indicates that values are atomic and do not need further evaluation and the axiom E-VAR specifies the semantics of a variable.

$$\eta \vdash v \Downarrow_e v \qquad \text{(E-VALUE)}$$

$$\frac{\eta(x) = v}{\eta \vdash x \Downarrow_e v} \qquad \text{(E-VAR)}$$

**Table 7: Evaluation rules for expressions.**

## B.3 Statements

In the semantics of our DSL the set $\mathcal{K}$ of configurations is defined as $\mathbf{Env} \times \mathbf{Object} \times \mathbf{Loc} \times \mathbf{PS} \times \mathbf{DAct} \times \mathbf{Stmt}$ for first-level statements and $\mathbf{Env} \times \mathbf{Object} \times \mathbf{Loc} \times \mathbf{PS} \times \mathbf{DAct} \times \mathbf{ActStmt}$ for second-level statements.

For first-level statements, we will represent specific configurations by both the meta-variable $k$ and by tuples of the form $\eta \mid os \mid s$, where: $\eta$ is the environment of variables, references and fresh identifiers; $os$ is the object set of the structured model; and $s$ is the statement to be evaluated.

For second-level statements, we will represent specific configurations by both the meta-variable $q$ and by tuples of the form $\eta \mid os \mid l \mid ps \mid as \mid s2$, where: $\eta$ and $os$ are as above; $ps$ is the set of properties of the object under the focus of the interpreter; $l$ points to the location in the model that currently receives the focus of the interpreter; $ps$ is the set of properties of the object under the focus of the interpreter; $as$ is the ordered set of deferred actions; and $s2$ is the statement to be evaluated. To enhance readability, we will abuse the notation so that whenever $as$ is not required, it will not be represented.

Given an input structured model $M$ represented by an or-dered set of objects $os \in \mathbf{ObjSet}$ and a first-level statement $s$, an initial configuration of first-level statements is defined as follows: the environment $\eta$ is initialized with the locations in $os$ for each object identifier and the map $new$ is also initialized for each class name that appears in the model.

*Definition 15.* Configuration with $M$ and $s$. A configuration with a model $M$ of model type $\mathcal{M}$ and a statement $s$ is a configuration $k = \eta \mid os \mid s$ where

$$\eta = VM \uplus loc(os) \uplus fresh(M)$$

where $VM \subset (\mathbf{Var} \rightharpoonup \mathcal{D}_b \uplus \mathbf{Var} \rightharpoonup \mathcal{O})$, $loc(M)$ obtains a map from the identifier of each object in $os$ to its location, and $fresh(M)$ obtains a map from class names in $\mathcal{M}_C$ to a fresh identifier.

Given an input structured model $M$ represented by an ordered set of objects $os \in \mathbf{ObjSet}$, an object identifier $o$ for the object under focus, and a second-level statement $s2$, an initial configuration of second-level statements is defined as follows:

*Definition 16.* Configuration with $M$, $o$ and $s$. A configuration with a model $M = (os, \Pi)$ of model type $\mathcal{M}$ and a second-level statement $s2$ is a configuration

$$q = \eta \mid os' \mid loc(o) \mid ps(\rho) \mid s2 \mid$$

where

$$\eta = VM \uplus loc(os) \uplus fresh(M)$$

where $VM \subset (\mathbf{Var} \rightharpoonup \mathcal{D}_b \uplus \mathbf{Var} \rightharpoonup \mathcal{O})$, $loc(M)$ obtains a map from the identifier of each object in $os$ to its location, $fresh(M)$ obtains a map from class names in $\mathcal{M}_C$ to a fresh identifier, and $\langle os'; \rho \rangle = os|_{loc(o)}$.

The set $\mathcal{K}_T$ of terminal configurations are defined as $\{(\eta \mid os \mid s) \in \mathcal{K} \mid s = ()\}$ for first-level statements and as $\{(\eta \mid os \mid l \mid ps \mid s2) \in \mathcal{K} \mid s2 = \bullet\}$ for second-level statements.

In our semantics, a transition is of the form

$$\eta \mid os \mid s \Downarrow_s \eta' \mid os' \mid s'$$

for first-level statements and of the form

$$\eta \mid os \mid l \mid ps \mid as \mid s2 \Downarrow_{s2} \eta' \mid os' \mid l' \mid ps' \mid as' \mid s2'$$

for second-level statements. The transition rules describing the semantics of our language are split in two subgroups, depending on whether they are used within a snapshot command or not. First-level statements, whose semantic rules are shown in Table 9, and correspond to the semantics of the creation of root objects, snapshots, declaration of variables, the first-level no-op, and sequences. Second-level statements, whose semantic rules shown in Table 8, correspond to statements that can only be used within a snapshot command. Second-level statements correspond to model actions, declaration of variables, the second-level no-op, and sequences.

We start by describing the semantic relation for second-level statements, which do not modify the set of objects $os$ in the configuration and their effects are applied to the property set that belongs to the object under focus (i.e. placed at location $l$).

- The model action $\texttt{create}(p,c)$, specified in rules E-Create and E-CreateBi, creates a new object of type $c$ with the fresh identifier $o$ and with a set of object properties. This set of properties is initialized to default values for the corresponding base type as produced by the operation $default(c)$. The location of the new object in the model is stored in the environment $\eta'$. The fresh object identifier of class $c$ is updated and stored in the environment $\eta'$.

Rule E-CreateBi deals with the case where the containment property $p$ is bidirectional by generating a deferred model action to set the opposite reference. The operations $noOp : \texttt{BRE} \to D_{Bool}$ and $getOp : \texttt{BRE} \rightharpoonup \texttt{BRE}$ are used to determine if a reference is bidirectional, i.e. it has an opposite end, and to get the opposite reference if it exists, respectively. These operations use the model components, from Definition 3, corresponding to opposite reference ends $\mathcal{M}_{oe}$ and to the transitive closure of the reified object subtype relation $\mathcal{M}_{sr}$ to find the opposite reference.

- The model action $\texttt{set}(p,e)$, specified by the rule E-SetAtt, assigns the value resulting from evaluating expression $e$ to property $p$ in the set of properties of the object under focus.

- The model action $\texttt{unset}(p)$, specified by the rule E-UnsetAtt, assigns the default value to property $p$ in the set of properties of the object under focus.

- The model action $\texttt{set}(p,x)$ adds the object reference $x$ to property $p$ in the set of properties of the object under focus. Rule E-RefUniSet applies when a reference is unidirectional, i.e. there is no opposite end in $\mathcal{M}_{oe}$ and rule E-RefBiSet applies when the reference is bidirectional.

Both rules are defined when the object to be referenced exists in the model, that is, if its object identifier is mapped to a location in the environment $\eta$. Both rules insert the object identifier at the end of the order set if the collection $is$ does not contain it, preserving the set semantics of the collection.

- The model action $\texttt{unset}(p,x)$ deletes the object reference $x$ from the set of identifiers contained by the property $p$ in the set of properties of the object under focus. Its semantics is specified by the rule E-RefUniUnset for unidirectional references and by the rule E-RefBiUnset for bidirectional references. In the second case, a new action is created in order to unset the opposite reference. Its application is deferred until the end of the execution of the containing snapshot statement.

- The model action $\texttt{setCmt}(p,x)$ takes the object from location $l'$ in model $\rho$, corresponding to the object pointed by $x$, and places it underneath the object under focus by adding it to the set of objects contained by property $p$. The new location of the object and of its contained objects is updated in the environment. The rule E-CmtUniSet specifies the semantics of the operation when the containment has no opposite reference and the rule E-CmtBiSet specifies the semantics of the model action when the containment has an opposite reference. In the second case, the opposite reference has to be updated in order to point to the new container object: if the object was a root object the reference to the container object was not set and it only needs to be set to the new container; otherwise the reference to the old container has to be removed as well.

- The model action unset($p, x$), specified by the rule E-CmtUnset, removes the object pointed by $x$ from the set of objects contained by the property $p$ in the set of properties of the object under focus. For an object $\rho$ in a model $\rho'$, the predicate $isolated(\rho, \rho')$ is satisfied when the object $\rho$ and its contained objects are not referenced anywhere else in the model $\rho'$. The environment is updated by removing those references corresponding to objects that are removed. This rule ensures that orphan nodes are not created by deleting a complete composite object (with its contained subobjects) when a containment is unset. Otherwise, we could create root objects that are not typed with root meta-classes. According to our notion of model type, such models are not well typed. The reason for forbidding this possibility is to respect Liskov's substitution principle when a model type is used as the type of the parameter in a model management operation, where only polymorphic types are allowed.

- In rule E-ActLet, a statement let $x = e$ in $s$ binds the value resulting from evaluating the expression $e$ to a new variable $x$ that is local to the block of statements $s$. Its execution does not alter the focus used in the last statement of $s$ and it preserves the effects of the statements $s$ on the structured model $\rho$. At the end of its execution, local variables created within the block of statements are forgotten but new or updated object variables are preserved, together with the locations of the new objects and the list of fresh identifiers for each class. The rule E-ActLetCreate specifies the creation of an object reference with the model action create($p, c$), creating a new object under the currently focussed object through containment $p$.

- The operator snapshot2 $x$ {$s2$} performs side effects on the model instance by replacing the property set, in the object identified by the reference $x$, with the property set that results from the application of the sequence of model actions $s2$. The execution specification of the statement snapshot2 $x$ {$s2$} in the rule E-ActSnapshot changes the focus of the interpreter to the object referenced by $x$ so that the intermediate execution steps contained in block $s2$ continue with this focus. This is achieved by updating the location $l$ and the property set $ps$ in the configuration of the intermediate evaluation transitions. As opposed to the first-level statement snapshot $x$ {$s2$}, the operator snapshot2 $x$ {$s2$} searches for the object to be manipulated inside the object under focus without having to change the operation context. This avoids traversals in the structured model every time it needs to be updated and all the effects are combined in the property set.

  At the end of the transition, the resulting changes are merged into the model by inserting at location $l'$, in the object that was originally under focus, a new object with the identifier $\rho'_o$ and the property set resulting from the execution of model actions in intermediate transitions with the expression $(<o' | ps> [<\rho'_o | ps'>]_{l'})_{ps}$. The focus is restored once the evaluation transition is completed.

- Finally, axiom E-FmaNext and rule E-FmaSeq enable the evaluation of sequences of statements as expected.

The semantic relation for first-level statements includes similar rules for the declaration of both first-level value variables (E-FmaLet) and first-level reference variables with (E-FmaLetCreate), the model actions (E-FmaCreateRoot) and (E-FmaDeleteRoot), the snapshot command (E-FmaSnapshot), the first-level no-op (E-FmaNext) and sequences (E-FmaSeq). The rules that are different from those presented for second-level statements are (E-FmaSnapshot), (E-FmaCreateRoot) and (E-FmaDeleteRoot).

The operator snapshot $x$ {$s2$} performs side effects on the model instance by replacing the property set, in the object identified by the reference $x$, with the property set that results from the action of a sequence of model manipulation statements. The execution specification of a statement in the snapshot $x$ {$s2$} in the rule E-FmaSnapshot changes the focus of the interpreter to the object pointed by $x$ so that the intermediate execution steps contained in block $s2$ continue with this focus. This is achieved by updating the location $l$ and the property set $ps$ in the configuration of the intermediate evaluation transitions. This avoids traversals in the structured model every time it needs to be updated and all the effects are combined in the property set. These effects are realized on the model when the interpreter finishes the evaluation of the containing snapshot statement as explained above. At the end of the transition, the new model instance $\rho'''$ is updated by inserting at location $l'$ the object $\rho''$ with the effects resulting from model actions in intermediate transitions, including those effects corresponding to deferred model actions. The focus is restored once the evaluation transition is completed. That is, the location under focus and the property set before the transition is performed are restored so that the interpreter can continue the execution of the program with the previous focus.

The rule (E-FmaCreateRoot) specifies how to create a root object when the interpreter has not been focussed on an object with the snapshot command. Apart from the fact that no containment is indicated in the action create($c$), the semantics is very similar to (E-FmaCreate). Similarly, the rule (E-FmaDeleteRoot) specifies how to delete a root object.

## C. PROPERTIES OF THE EVALUATION RELATION

The following theorems tells us that the specified semantics provides a unique interpretation for each DSL statement. We first prove the determinacy of the evaluation relation $\Downarrow_{s2}$ and then articulate the determinacy of the evaluation relation $\Downarrow_s$.

*Lemma 3.* (Determinacy of the evaluation relation $\Downarrow_{s2}$) If $q \Downarrow_{s2} q'$ and $q \Downarrow_{s2} q''$, then $q' = q''$.

PROOF. The proof follows by induction on the structure of the transition $q \Downarrow_{s2} q'$.

Base cases are resolved by analysing the overlaps of the configurations $q$ in the axioms of the inference system. Given that the statement $s$ in $q$ cancels the conflicts, we only consider the cases where the statement is the same. Whenever the last rule of the transition $q \Downarrow q''$ coincides with the rule of first transtion, by direct proof we can check $q' = q''$. This leads us to the cases formed by the pair E-RefUniSet and E-CmtUniSet, the pair E-RefUniUnset and E-CmtUniUnset, the pair E-RefBiSet and E-CmtBiSet, the pair E-RefBiUnset and E-CmtBiUnset. In all cases,

$$\frac{\begin{array}{c} noOp(bCE(last(l), p)) \\ o = \eta(c) \quad \eta' = \eta[c \mapsto fresh(o); o \mapsto l.p.o] \quad \rho' = \texttt{<}o\,|\,default(c)\texttt{>} \end{array}}{\eta \,|\, os \,|\, l \,|\, p = os'', ps \,|\, as \,|\, \texttt{create}(p, c) \Downarrow_{s2} \eta' \,|\, os \,|\, l \,|\, p = os''\,\rho', ps \,|\, as \,|\, \bullet} \quad \text{(E-Create)}$$

$$\frac{\begin{array}{c} bRE(c', p') = getOp(bRE(last(l), p)) \\ o = \eta(c) \quad \eta' = \eta[c \mapsto fresh(o); o \mapsto l.p.o] \quad \rho' = \texttt{<}o\,|\,default(c)\texttt{>} \quad a = set(o, p', last(l)) \end{array}}{\eta \,|\, os \,|\, l \,|\, p = os'', ps \,|\, as \,|\, \texttt{create}(p, c) \Downarrow_{s2} \eta' \,|\, os \,|\, l \,|\, p = os''\,\rho', ps \,|\, as\, a \,|\, \bullet} \quad \text{(E-CreateBi)}$$

$$\frac{\eta \vdash e \Downarrow_e v'}{\eta \,|\, os \,|\, l \,|\, p = v, ps \,|\, as \,|\, \texttt{set}(p, e) \Downarrow_{s2} \eta \,|\, os \,|\, l \,|\, p = v', ps \,|\, as \,|\, \bullet} \quad \text{(E-AttSet)}$$

$$\frac{}{\eta \,|\, os \,|\, l \,|\, p = v, ps \,|\, as \,|\, \texttt{unset}(p) \Downarrow_{s2} \eta \,|\, os \,|\, l \,|\, p = default(v), ps \,|\, as \,|\, \bullet} \quad \text{(E-AttUnset)}$$

$$\frac{noOp(bRE(last(l), p)) \quad \eta \vdash x \Downarrow_e o \quad l' = \eta(o)}{\eta \,|\, os \,|\, l \,|\, p = is, ps \,|\, as \,|\, \texttt{set}(p, x) \Downarrow_{s2} \eta \,|\, os \,|\, l \,|\, p = (o \in is)?\, is : is\, o, ps \,|\, as \,|\, \bullet} \quad \text{(E-RefUniSet)}$$

$$\frac{\begin{array}{c} bRE(c', p') = getOp(bRE(last(l), p)) \\ \eta \vdash x \Downarrow_e o \quad l' = \eta(o) \quad is' = (o \in is)?\, is : is\, o \quad a = set(o, p', last(l)) \end{array}}{\eta \,|\, os \,|\, l \,|\, p = is, ps \,|\, as \,|\, \texttt{set}(p, x) \Downarrow_{s2} \eta \,|\, os \,|\, l \,|\, p = is' \,|\, (o \in is)?\, as : as\, a \,|\, \bullet} \quad \text{(E-RefBiSet)}$$

$$\frac{noOp(bRE(last(l), p)) \quad \eta \vdash x \Downarrow_e o}{\eta \,|\, os \,|\, l \,|\, p = o\, is, ps \,|\, as \,|\, \texttt{unset}(p, x) \Downarrow_{s2} \eta \,|\, os \,|\, l \,|\, p = is, ps \,|\, as \,|\, \bullet} \quad \text{(E-RefUniUnset)}$$

$$\frac{bRE(c', p') = getOp(bRE(last(l), p)) \quad \eta \vdash x \Downarrow_e o \quad a = unset(o, p', last(l))}{\eta \,|\, \rho \,|\, l \,|\, p = o\, is, ps \,|\, as \,|\, \texttt{unset}(p, x) \Downarrow_{s2} \eta \,|\, \rho \,|\, l \,|\, p = is, ps \,|\, as\, a \,|\, \bullet} \quad \text{(E-RefBiUnset)}$$

$$\frac{noOp(bRE(last(l), p)) \quad \eta \vdash x \Downarrow_e o \quad l' = \eta(o) \quad \langle os', \rho'' \rangle = os|_{l'} \quad \eta'' = \eta[o \mapsto l.p.o]}{\eta \,|\, os \,|\, l \,|\, p = os'', ps \,|\, as \,|\, \texttt{set}(p, x) \Downarrow_{s2} \eta'' \,|\, os' \,|\, l \,|\, p = ((\rho'' \in os'')?\, os'' : os''\, \rho''), ps \,|\, as \,|\, \bullet} \quad \text{(E-CmtUniSet)}$$

$$\frac{\begin{array}{c} bRE(c', p') = getOp(bCE(last(l), p)) \quad \eta \vdash x \Downarrow_e o \quad o' = container(\eta(o)) \quad \langle os', \rho'' \rangle = os|_{\eta(o)} \\ \eta'' = \eta[o \mapsto l.p.o] \quad as' = (containsRoot(o'))?\, set(o, p', last(l))\, as : set(o, p', last(l))\, unset(o, p', o')\, as \end{array}}{\eta \,|\, os \,|\, l \,|\, p = os'', ps \,|\, as \,|\, \texttt{set}(p, x) \Downarrow_{s2} \eta'' \,|\, os' \,|\, l \,|\, p = ((\rho'' \in os'')?\, os'' : os''\, \rho''), ps \,|\, (\rho'' \in os'')?\, as : as' \,|\, \bullet} \quad \text{(E-CmtBiSet)}$$

$$\frac{isolated(\texttt{<}o\,|\,ps''\texttt{>}, \rho) \quad \eta \vdash x \Downarrow_e o \quad \eta' = \eta|_{\overline{oids(\texttt{<}o\,|\,ps''\texttt{>})}}}{\eta \,|\, os'' \,|\, l \,|\, p = \texttt{<}o\,|\,ps''\texttt{>}\, os, ps \,|\, as \,|\, \texttt{unset}(p, x) \Downarrow_{s2} \eta' \,|\, os'' \,|\, l \,|\, p = os, ps \,|\, as \,|\, \bullet} \quad \text{(E-CmtUnset)}$$

$$\frac{\eta \vdash e \Downarrow_e v \quad \eta[x \mapsto v] \,|\, \rho \,|\, l \,|\, ps \,|\, as \,|\, s \Downarrow_{s2} \eta' \,|\, \rho' \,|\, l' \,|\, ps' \,|\, as' \,|\, \bullet \quad \eta'' = \eta \cup (\eta' \backslash (\eta'|_{\mathbf{Var}}))}{\eta \,|\, \rho \,|\, l \,|\, ps \,|\, as \,|\, \texttt{let } x = e \texttt{ in } s2 \Downarrow_{s2} \eta'' \,|\, \rho' \,|\, l' \,|\, ps' \,|\, as' \,|\, \bullet} \quad \text{(E-ActLet)}$$

$$\frac{\begin{array}{c} \eta \,|\, \rho \,|\, l \,|\, ps \,|\, as \,|\, \texttt{create}(p, c) \Downarrow_{s2} \eta' \,|\, \rho \,|\, l \,|\, ps' \,|\, as' \,|\, \bullet \\ (o \mapsto l) = \eta' \backslash \eta \quad \eta'[x \mapsto o] \,|\, \rho \,|\, l \,|\, ps' \,|\, as' \,|\, s2 \Downarrow_{s2} \eta'' \,|\, \rho' \,|\, l' \,|\, ps'' \,|\, as'' \,|\, \bullet \quad \eta'' = \eta \cup (\eta' \backslash (\eta'|_{\mathbf{Var}})) \end{array}}{\eta \,|\, \rho \,|\, l \,|\, ps \,|\, as \,|\, \texttt{let } x = \texttt{create}(p, c) \texttt{ in } s2 \Downarrow_{s2} \eta'' \,|\, \rho' \,|\, l' \,|\, ps'' \,|\, as'' \,|\, \bullet} \quad \text{(E-ActLetCreate)}$$

$$\frac{\eta \vdash x \Downarrow_e o \quad l' = \eta(o) \quad o' = last(l) \quad \langle os', \rho' \rangle = \texttt{<}o'\,|\,ps\texttt{>}|_{l'} \quad \eta \,|\, os \,|\, l' \,|\, \rho'_{ps} \,|\, as \,|\, s2 \Downarrow_{s2} \eta' \,|\, os' \,|\, l'' \,|\, ps' \,|\, as' \,|\, \bullet}{\eta \,|\, os \,|\, l \,|\, ps \,|\, as \,|\, \texttt{snapshot2 } x\, \{s2\} \Downarrow_{s2} \eta' \,|\, os' \,|\, l \,|\, (\texttt{<}o'\,|\,ps\texttt{>}[\texttt{<}\rho'_o\,|\,ps'\texttt{>}]_{l'})_{ps} \,|\, as' \,|\, \bullet} \quad \text{(E-ActSnapshot)}$$

$$\frac{\eta \,|\, \rho \,|\, l \,|\, ps \,|\, as \,|\, s2 \Downarrow_{s2} \eta' \,|\, \rho' \,|\, l' \,|\, ps' \,|\, as \,|\, \bullet}{\eta \,|\, \rho \,|\, l \,|\, ps \,|\, as \,|\, \bullet;\, s2 \Downarrow_{s2} \eta' \,|\, \rho' \,|\, l' \,|\, ps' \,|\, as \,|\, \bullet} \quad \text{(E-ActNext)}$$

$$\frac{\eta \,|\, \rho \,|\, l \,|\, ps \,|\, as \,|\, s2_1 \Downarrow_{s2} \eta' \,|\, \rho' \,|\, l' \,|\, ps' \,|\, as \,|\, \bullet \quad \eta' \,|\, \rho' \,|\, l' \,|\, ps' \,|\, as \,|\, s2_2 \Downarrow_{s2} \eta'' \,|\, \rho'' \,|\, l'' \,|\, ps'' \,|\, as \,|\, \bullet}{\eta \,|\, \rho \,|\, l \,|\, ps \,|\, as \,|\, s2_1;\, s2_2 \Downarrow_{s2} \eta'' \,|\, \rho'' \,|\, l'' \,|\, ps'' \,|\, as \,|\, \bullet} \quad \text{(E-ActSeq)}$$

Table 8: SOS of second-level statements.

$$\frac{o = \eta(c) \quad \eta' = \eta[c \mapsto fresh(o); o \mapsto o] \quad \rho' = \mathord{<}o\,|\,default(c)\mathord{>}}{\eta\,|\,os\,|\,\texttt{create}(c) \Downarrow_s \eta'\,|\,os\,\rho'\,|\,()} \quad \text{(E-FmaCreateRoot)}$$

$$\frac{\eta \vdash x \Downarrow_e o \quad isolated(\rho, os)}{\eta\,|\,\mathord{<}o\,|\,ps\mathord{>}\,os\,|\,\texttt{delete}(x) \Downarrow_s \eta\,|\,os\,|\,()} \quad \text{(E-FmaDeleteRoot)}$$

$$\frac{\eta \vdash e \Downarrow_e v \quad \eta[x \mapsto v]\,|\,os\,|\,s \Downarrow_s \eta'\,|\,os'\,|\,() \quad \eta'' = \eta \cup (\eta' \backslash (\eta'|_{\mathbf{Var}}))}{\eta\,|\,os\,|\,\texttt{let } x = e \texttt{ in } s \Downarrow_s \eta''\,|\,os'\,|\,()} \quad \text{(E-FmaLet)}$$

$$\frac{\eta\,|\,os\,|\,\texttt{create}(c) \Downarrow_s \eta'\,|\,os'\,|\,() \quad (o \mapsto l) = \eta' \backslash \eta \quad \eta'[x \mapsto o]\,|\,os'\,|\,s \Downarrow_s \eta''\,|\,os''\,|\,() \quad \eta''' = \eta \cup (\eta'' \backslash (\eta''|_{\mathbf{Var}}))}{\eta\,|\,os\,|\,\texttt{let } x = \texttt{create}(c) \texttt{ in } s \Downarrow_s \eta'''\,|\,os''\,|\,()} \quad \text{(E-FmaLetCreate)}$$

$$\frac{\eta \vdash x \Downarrow_e o \quad l' = \eta(o) \quad \langle os', \rho'' \rangle = os|_{l'} \quad \eta\,|\,os'\,|\,l'\,|\,\rho''_{ps}\,|\,\emptyset\,|\,s2 \Downarrow_{s2} \eta'\,|\,os'''\,|\,l''\,|\,ps'\,|\,as\,|\,\bullet}{\eta\,|\,os\,|\,\texttt{snapshot } x\,\{s2\} \Downarrow_s \eta'\,|\,eval(as, \eta', os'''[\mathord{<}\rho''_o\,|\,ps'\mathord{>}]_{l'})\,|\,()} \quad \text{(E-FmaSnapshot)}$$

$$\frac{\eta\,|\,\rho\,|\,s \Downarrow_s \eta'\,|\,\rho'\,|\,()}{\eta\,|\,\rho\,|\,();s \Downarrow_s \eta'\,|\,\rho'\,|\,()} \quad \text{(E-FmaNext)}$$

$$\frac{\eta\,|\,\rho\,|\,s_1 \Downarrow_s \eta'\,|\,\rho'\,|\,() \quad \eta'\,|\,\rho'\,|\,s_2 \Downarrow_s \eta''\,|\,\rho''\,|\,()}{\eta\,|\,\rho\,|\,s_1;s_2 \Downarrow_s \eta''\,|\,\rho''\,|\,()} \quad \text{(E-FmaSeq)}$$

Table 9: SOS of first-level statements.

$p$ may refer to either a property bound to a set object identifiers (in E-RefUniSet, E-RefUniUnset, E-RefBiSet and E-RefBiUnset) or to a property bound to a set of objects (E-CmtUniSet, E-CmtUniUnset, E-CmtBiSet and E-CmtBiUnset). If we assume that transitions E-RefUniSet and E-CmtUniSet are applied over the same configuration $q$, two different properties with the same name $p$ and with different values must exist in the set of properties of the object under focus. However, this is not possible since a property name $p$ uniquely identifies a property in an object by assumption. The same reasoning applies to the other pairs of transitions.

For proving the induction step, we assume the desired property for all intermediate transitions, and proceed by case analysis of the evaluation rule used in the last transition. If the last rule used in the transition $q \Downarrow q'$ is E-ActLet, $q$ has the form $\eta\,|\,\rho\,|\,l\,|\,ps\,|\,as\,|\,\texttt{let } x = e \texttt{ in } s$ where $\eta \vdash e \Downarrow_e v$ and $\eta[x \mapsto v]\,|\,\rho\,|\,l\,|\,ps\,|\,as\,|\,s \Downarrow \eta'\,|\,\rho'\,|\,l'\,|\,ps'\,|\,as'\,|\,()$ for some $\eta', \rho', l', ps', as'$. The last rule in the transition $q \Downarrow q''$ can only be E-ActLet for the same reasons stated above. Therefore, $q$ has the form $\eta\,|\,\rho\,|\,l\,|\,ps\,|\,as\,|\,\texttt{let } x = e \texttt{ in } s$ where $\eta \vdash e \Downarrow_e v$ and $\eta[x \mapsto v]\,|\,\rho\,|\,l\,|\,ps\,|\,as\,|\,s \Downarrow \eta''\,|\,\rho''\,|\,l''\,|\,ps''\,|\,as''\,|\,()$ for some $\eta'', \rho'', l'', ps''$. Since the transitions $\eta[x \mapsto v]\,|\,\rho\,|\,l\,|\,ps\,|\,as\,|\,s \Downarrow \eta'\,|\,\rho'\,|\,l'\,|\,ps'\,|\,as'\,|\,()$ and $\eta[x \mapsto v]\,|\,\rho\,|\,l\,|\,ps\,|\,as\,|\,s \Downarrow \eta''\,|\,\rho''\,|\,l''\,|\,ps''\,|\,as''\,|\,()$ are intermediate transitions of the root transitions $q \Downarrow q'$ and $q \Downarrow q''$, resp., by the induction hypothesis we obtain that $\eta'\,|\,\rho'\,|\,l'\,|\,ps'\,|\,as'\,|\,() = \eta''\,|\,\rho''\,|\,l''\,|\,ps''\,|\,as''\,|\,()$ yielding $q' = q''$. The cases for rules E-ActLetCreate, E-ActSnapshot, E-ActNext and E-ActSeq follow in a similar way. In the case of E-ActLetCreate, determinacy is ensured up to object identifier renaming, given that the order in which an object is created determines its identifier. □

*Theorem 4.* (Determinacy of the evaluation relation $\Downarrow_s$) If $k \Downarrow_s k'$ and $k \Downarrow_s k''$, then $k' = k''$.

Proof. (Idea.) The proof is predicated by structural induction on $\Downarrow_s$ following a similar reasoning used in the proof of Lemma 3 and by using Lemma 3 for the statement `snapshot`. □

To prove the termination of the evaluation relations $\Downarrow_s$ and $\Downarrow_{s2}$, we define an embedding into $(\mathbb{N}, >)$ by using an overloaded measure function $\xi$ to compute the size of configurations, statements and expressions.

*Definition 17.* (Measure function) The measure function $\xi$ is defined over configurations, statements and expressions as follows:

$$\xi(\eta\,|\,\rho\,|\,s) = \xi(s)$$
$$\xi(\texttt{create}(c)) = \xi(\texttt{delete}(x)) = \xi(()) = 1$$
$$\xi(\texttt{let } x = e \texttt{ in } s) = 1 + \xi(e) + \xi(s)$$
$$\xi(\texttt{let } x = \texttt{create}(c) \texttt{ in } s) = 1 + \xi(e) + \xi(s)$$
$$\xi(\texttt{snapshot } x\,\{s2\}) = 1 + \xi(s2)$$
$$\xi(s; s') = 1 + \xi(s) + \xi(s')$$
$$\xi(\eta\,|\,\rho\,|\,l\,|\,ps\,|\,as\,|\,s2) = \xi(s2)$$
$$\xi(\texttt{create}(p, c)) = \xi(\texttt{unset}(p))$$
$$= \xi(\texttt{unset}(p, x)) = \xi(\bullet) = 1$$
$$\xi(\texttt{set}(p, e)) = 1 + \xi(e)$$
$$\xi(\texttt{let } x = e \texttt{ in } s2) = 1 + \xi(e) + \xi(s2)$$
$$\xi(\texttt{let } x = \texttt{create}(p, c) \texttt{ in } s2) = 2 + \xi(s2)$$
$$\xi(\texttt{snapshot2 } x\,\{s2\}) = 1 + \xi(s2)$$
$$\xi(s2; s2') = 1 + \xi(s2) + \xi(s2')$$
$$\xi(x) = \xi(v) = 1$$

*Lemma 4.* (Measure function $\xi$ is monotone) The measure function $\xi$ is a monotone mapping, i.e. for each evaluation transition $k \Downarrow_s k'$, $q \Downarrow_s q'$ and $e \Downarrow_e e'$, we have that $\xi(k) > \xi(k')$.

Proof. The proof is by induction on the structure of an evaluation transition $k \Downarrow k'$. We show it for the evaluation relation $\Downarrow_s$. For each transition, we check that $\xi$ is a monotone mapping between $(\mathbf{Stmt}, \Downarrow_s)$ and $(\mathbb{N}, >)$. This check is immediate for each evaluation rule, but for when the last applied rule is E-FmaSeq, in which case we have that $k = \eta\,|\,\rho\,|\,s_1;s_2$ and that $k' = \eta\,|\,\rho\,|\,s_1';s_2$. Therefore,

$\xi(k) = \xi(s_1; s_2) = \xi(s_1) + \xi(s_2)$ and $\xi(k') = \xi(s_1'; s_2) = \xi(s_1') + \xi(s_2)$. By the induction hypothesis, we obtain that $\xi(s_1) > \xi(s_1')$. Hence, $\xi(k) > \xi(k')$. The proofs for $\Downarrow_{s2}$ and $\Downarrow_e$ proceed similarly. □

*Definition 18.* (Configuration in normal form) A configuration $k$ is said to be in normal form if there is no $k'$ such that $k \Downarrow k'$.

*Theorem 5.* (Termination of $\Downarrow_s$) For each configuration $k$, either it is in normal form or there is a configuration $k'$ in normal form such that $k \Downarrow k'$.

PROOF. The first case is trivial. For the second case, we consider that $\xi$ is a monotone embedding from $(\mathbf{Stmt}, \Downarrow_s)$ into $(\mathbb{N}, >)$ by Lemma 4. By the inverse image construction we have that $\Downarrow \subseteq \xi^{-1}(>)$. Since $(\mathbb{N}, >)$ is known to be terminating, then $(\mathbf{Stmt}, \Downarrow_s)$ is also terminating. □

## D. VALIDITY THEOREM

In this section, we show that the SOS specification of FMA characterizes programs that are safe with respect to the notion of valid model, i.e. the validity of model transformations. In words, given a valid model, the execution of a FMA program will always result in a valid model, or else it will return a trapped error. In addition, the semantics of FMA is deterministic and terminating as shown in [10, Appendix C].

In what follows, we augment the evaluation relation with error detection. Then we provide the scaffolding required to prove that FMA is a safe language with respect to valid models, which amounts to show its totality and the subject reduction property of the evaluation relation. That is, that execution cannot get stuck due to an untrapped error and that all models that can be computed are valid.

### D.1 Error Detection and Propagation

The evaluation relation is augmented with rules for making explicit trapped errors that invalidate a model and with rules for propagating those errors. An error is displayed as a special statement error that halts the execution. Table 10 and Table 11 show the list of rules for identifying and for propagating errors. For each rule, we indicate the SOS rule on which it is based, the type of rule (I if the rule identifies a trapped error and P if the rule propagates the error), the condition that enables the rule (i.e. the precondition that differentiates the error rule from its SOS counterpart), and the type of error causing an invalid model that it detects.

### D.2 Totality of Evaluation Relation

We introduce the notion of FMA closed programs based on the set $FV(s)$ of free variables for a FMA program $s$:

- for first-level statements:

$$FV(\texttt{snapshot } x \; \{s2\}) = FV(s2)$$
$$FV(\texttt{let } x = e \texttt{ in } s) = FV(e) - FV(s)$$
$$FV(\texttt{let } x = \texttt{create}(c) \texttt{ in } s) = FV(e) - FV(s)$$
$$FV(\texttt{create}(c)) = \emptyset$$
$$FV(s; s') = FV(s) \cup FV(s')$$
$$FV(()) = \emptyset$$

- for expressions:

$$FV(v) = \emptyset$$
$$FV(x) = \{x\}$$

- for second-level statements:

$$FV(\texttt{snapshot2 } x \; \{s2\}) = FV(s2)$$
$$FV(\texttt{let } x = e \texttt{ in } s2) = FV(e) \cup (FV(s2) - \{x\})$$
$$FV(\texttt{let } x = \texttt{create}(c) \texttt{ in } s2) = FV(s2) - \{x\}$$
$$FV(\texttt{create}(p, c)) = \emptyset$$
$$FV(\texttt{set}(p, x)) = \{x\}$$
$$FV(\texttt{setCmt}(p, x)) = \{x\}$$
$$FV(\texttt{unset}(p)) = \emptyset$$
$$FV(\texttt{unset}(p, x)) = \{x\}$$
$$FV(s2; s2') = FV(s2) \cup FV(s2')$$
$$FV(\bullet) = \emptyset$$

*Definition 19.* Free variables of a FMA program. Given a FMA program $s \in \mathbf{Stmt}$, its set of free variables is defined as $FV(s)$.

A closed FMA program is a program where all contained free variables are bound, i.e. where there are no free variables. More formally:

*Definition 20.* Closed FMA program. A FMA program $s \in \mathbf{Stmt}$ is closed iff $FV(s) = \emptyset$.

*Definition 21.* Configuration with $M$ and $s$. A configuration with a model $M$ of model type $\mathcal{M}$ and a statement $s$ is a configuration $k = \eta \,|\, M \,|\, s$ where

$$\eta = VM \uplus loc(M) \uplus fresh(M)$$

where where $VM \subset (\mathbf{Var} \rightharpoonup \mathcal{D}_b \uplus \mathbf{Var} \rightharpoonup \mathcal{O})$, $loc(M)$ obtains a map from the identifier of each object in $M$ to its location, and $fresh(M)$ obtains a map from class names in $\mathcal{M}_C$ to a fresh identifier.

*Definition 22.* Configuration in normal form. A configuration $k$ is in normal form iff there is no configuration $k'$ such that $k \Downarrow_s k'$.

PROOF. It is easy to check that none of the rules belonging either to the evaluation relation or to the error relation for first-level statements can be applied. □

*Lemma 5.* Canonical forms for $\Downarrow_s$. Given a configuration $k$, if $k_s = ()$ then $k$ is in normal form with respect to $\Downarrow_s$.

PROOF. It is straightforward to check that none of the rules belonging either to the evaluation relation or to the error relation for second-level statements can be applied. □

*Theorem 6.* Totality of the evaluation relation $\Downarrow_s$. Given a configuration $k$ with model $M$ and closed FMA program $s$, where $M : \mathcal{M}_{pre}$ and a configuration $k'$, if $k \Downarrow_s k'$ then $k$ is in normal form, in which case $k_s' = ()$ or $k_s' = \texttt{error}$.

The proof requires a lemma stating a similar result for second-level statements.

A second-level statement $s2$ in a configuration $q$ is closed if all its free variables are bound in the variable environment $\eta$ in $q$. A second-level configuration with a model $M$ and a closed second-level statement $s2$, in addition requires a location and the property set of the object under focus.

| Error rule | SOS rule | type (I/P) | precondition | error |
|---|---|---|---|---|
| E-FmaDeleteRootErrorIso | E-FmaDeleteRoot | I | object not isolated | dangling edge |
| E-FmaDeleteRootErrorRef | E-FmaDeleteRoot | I | reference not mapped to a root location | dangling edge |
| E-FmaLetStmtError | E-FmaLet | P | – | – |
| E-FmaLetExprError | E-FmaLet | P | – | when the error comes from the evaluation of the expression |
| E-FmaLetCreateError | E-FmaLetCreate | P | – | – |
| E-FmaSnapshotErrorRef | E-FmaSnapshot | I | reference not mapped to a location | dangling reference |
| E-FmaSnapshotErrorProp | E-FmaSnapshot | P | – | – |
| E-FmaNextErrorProp | E-FmaNext | P | – | – |
| E-FmaSeqErrorProp1, E-FmaSeqErrorProp2 | E-FmaSeq | P | – | – |

Table 10: Error rules for first-level statements.

| Error rule | SOS rule | type (I/P) | precondition | error |
|---|---|---|---|---|
| E-RefSetError | E-RefUniSet, E-RefBiSet | I | reference not mapped to a location | dangling reference |
| E-RefSetErrorCmt | E-RefBiSet, | I | cannot update opposite to containment | opposite to containment |
| E-RefUnsetError1 | E-RefUniUnset, E-RefBiUnset | I | reference not mapped to a location | dangling reference |
| E-RefUnsetError2 | E-CmtUnset | I | reference not found in object under focus | dangling reference |
| E-RefUnsetErrorCmt | E-RefBiUnset, | I | cannot update opposite to containment | opposite to containment |
| E-CmtSetError | E-CmtUniSet, E-CmtBiSet | I | reference not mapped to a location | dangling reference |
| E-CmtSetErrorHoist | E-CmtUniSet, E-CmtBiSet | I | an object contained by the object under focus cannot be hoisted | dangling reference |
| E-CmtUnsetErrorRef1 | E-CmtUnset | I | reference not mapped to a location | dangling reference |
| E-CmtUnsetErrorRef2 | E-CmtUnset | I | referenced object is not a child of the object under focus | dangling reference |
| E-CmtUnsetErrorIso | E-CmtUnset | I | object not isolated | dangling reference |
| E-ActSnapshotErrorRef | E-ActSnapshot | I | reference not mapped to a location | dangling reference |
| E-ActSnapshotErrorProp | E-ActSnapshot | P | – | – |
| E-ActLetStmtErrorProp | E-ActLet | P | error occurs in statement | – |
| E-ActLetExprErrorProp | E-ActLet | P | error occurs in expression | – |
| E-ActLetCreateError | E-ActLetCreate | P | – | – |
| E-ActNextErrorProp | E-ActNext | P | – | – |
| E-ActSeqErrorProp1, E-ActSeqErrorProp2 | E-ActSeq | P | – | – |

Table 11: Error rules for second-level statements.

*Lemma 6.* Canonical forms for $\Downarrow_{s2}$. Given a configuration $k$, if $k_s = \bullet$ then $k$ is in normal form with respect to $\Downarrow_{s2}$.

*Lemma 7.* Totality of the evaluation relation $\Downarrow_{s2}$. Given a configuration $q$ with a valid model $M$ and a closed FMA statement $s2$, and a configuration $q'$, if $q \Downarrow_{s2} q'$ then $q'$ is in normal form, and either $q_{s2} = \bullet$ or $q_s = \mathtt{error}$.

PROOF. By induction on the structure of second-level FMA statements. The base cases correspond to the terms:

- statement $\mathtt{create}(p, c)$: we can only apply (E-CREATE) and get $q'' = \bullet$.
- statement $\mathtt{set}(p, x)$:
  – with rules (E-ATTSET), (E-REFUNISET), (E-REFBISET) we get $q'' = \bullet$;
  – with rule (E-REFSETERROR), we get $q'' = \mathtt{error}$.
- statement $\mathtt{setCmt}(p, x)$
  – with rules (E-CMTUNISET) and (E-CMTBISET), we get $q'' = \bullet$;
  – with rule (E-CMTSETERROR), (E-CMTSETERRORHOIST), we get $q'' = \mathtt{error}$.
- statement $\mathtt{unset}(p)$: with rule (E-ATTUNSET), we get $q'' = \bullet$.
- statement $\mathtt{unset}(p, x)$:
  – (E-REFUNIUNSET), (E-REFBIUNSET), (E-CMTUNSET), we get $q'' = \bullet$;
  – (E-REFUNSETERROR), (E-CMTUNSETERRORREF), (E-CMTUNSETERRORISO), we get $q'' = \mathtt{error}$;
- statement $\bullet$: normal form by lemma 6

The induction step cases correspond to

- statement $\mathtt{snapshot2}\ x\ \{s2\}$: by using the induction hypothesis, it follows that $q'' = \bullet$ with rule (E-ACTSNAPSHOT) and that $q'' = \mathtt{error}$ with rule (E-ACTSNAPSHOTERROR).
- statement $\mathtt{let}\ x = e\ \mathtt{in}\ s2$: by using the induction hypothesis, it follows that $q'' = \bullet$ with rule (E-LET) and that $q'' = \mathtt{error}$ with rule (E-LETERROR).
- statement $\mathtt{let}\ x = \mathtt{create}(c)\ \mathtt{in}\ s2$: by using the induction hypothesis, it follows that $q'' = \bullet$ with rule (E-LETCREATE) and that $q'' = \mathtt{error}$ with rule (E-LETCREATEERROR).
- statement $s2; s2$: by using the induction hypothesis, it follows that $q'' = \bullet$ with rules (E-ACTNEXT) and (E-ACTSEQ), and that $q'' = \mathtt{error}$ with rules (E-ACTNEXTERROR) and (E-ACTSEQERROR).

□

PROOF. Theorem 6 (Totality of the evaluation relation) By induction on the structure of the FMA programs, show that for each first-level configuration with model $M$ and a statement that is different from () we can apply a SOS rule that leads to the desired result.

The base cases correspond to the terms

- statement $\mathtt{create}(c)$: with rule (E-FMACREATEROOT), we get $k = ()$;
- statement $\mathtt{delete}(x)$: we get $k = ()$ with rule (E-FMACREATEROOT) and we get $k = \mathtt{error}$ with rules (E-FMADELETEROOTERRORISO) and (E-FMADELETEROOTERRORREF);

- $\mathtt{snapshot}\ x\ \{s2\}$: by using lemma 7 for showing the totality of the evaluation relation for $s2$, we get $k = ()$ with rule (E-SNAPSHOT), and we get $k = \mathtt{error}$ with rule (E-SNAPSHOTERROR).
- statement (): is in normal form by Lemma 5.

The induction step cases correspond to the statements $\mathtt{let}\ x = e\ \mathtt{in}\ s$, $\mathtt{let}\ x = \mathtt{create}(c)\ \mathtt{in}\ s$ and $s; s$ are proved as in the proof given for Lemma 7.

□

## D.3 Validity of Evaluation Relation

*Theorem 7.* $\Downarrow_s$ preserves model validity. Given a configuration $k$ with model $M$ and a closed FMA program $s$, and a configuration $k'$ with model $M'$, if $M : \mathcal{M}_{pre}$ and $k \Downarrow_s k'$ then $M' : \mathcal{M}_{pre}$.

*Lemma 8.* Given $M = (os, \Pi)$ such that $M : \mathcal{M}_{pre}$, a configuration $k$ with $M$, where $k = \eta\,|\,os\,|\,s$ for a closed FMA program $s$, and a second level statement $s2$ closed with respect to $\eta$, if $< os'; \rho > = os|_l$ and

$$\eta\,|\,os'\,|\,l\,|\,\rho_{ps}\,|\,\emptyset\,|\,s2 \Downarrow_{s2} \eta'\,|\,os'\,|\,l''\,|\,ps'\,|\,as\,|\,\bullet$$

then $eval(as, \eta', <\rho_o\,|\,ps'>[os']_l) : \mathcal{M}_{pre}$.

PROOF. By induction on the structure of second-level FMA statements, we show that the execution of statements do not violate the two defining conditions of the notion of valid (structured) model. Namely, that:

1. for any reference $p = ref(o\ is)$, there is an object in the model with object identifier $o$, which is ensured if $q_\eta(o)$ is defined.

2. for any two opposite references $p$ in meta-class $c$ and $p'$ in meta-class $c'$, $bRE(c', p') = \mathcal{M}_{oe}(bRE(c, p))$, including the case where the opposite reference is a containment $bRE(c', p') = \mathcal{M}_{oe}(bCE(c, p))$ or $bCE(c', p') = \mathcal{M}_{oe}(bRE(c, p))$, if both an object with identifier $o$ points to an object with identifier $o'$ through reference $p$ and the object with identifier $o'$ points to the object with identifier $o$ through reference $p'$.

The base cases correspond to:

- statement $\mathtt{create}(p, c)$: we can only apply (E-CREATE) and the statements are proved vacuously.
- statement $\mathtt{set}(p, x)$:
  – with rules (E-ATTSET), the conditions *1)* and *2)* are vacuously proved.
  – with (E-REFUNISET), a reference is added only if $q_\eta(o)$ is defined. So adding the reference satisfies condition *1)*. Condition *2)* is proved vacuously.
  – with (E-REFBISET), a reference is added only if $q_\eta(o)$ and condition *1)* is satisfied as above. In addition, (E-REFBISET) creates a deferred action to set the opposite direction of the reference $p$ in the object referenced by $x$, ensuring that condition *2)* is satisfied. The deferred action is applied by the *eval* operation.
- statement $\mathtt{setCmt}(p, x)$ is proved similarly with the rules (E-CMTUNISET) and (E-CMTBISET).
- statement $\mathtt{unset}(p)$: rule (E-ATTUNSET) satisfies the conditions *1)* and *2)* vacuously.

- statement unset($p, x$):
  - rules (E-RefUniUnset) and (E-CmtUnset) satisfy the conditions 1) and 2) vacuously. In the second rule, when a contained object is removed, its contents are also removed, including references.
  - rule (E-RefBiUnset), creates a deferred action to unset the opposite direction of the reference $p$ in the object referenced by $x$, ensuring that condition 2) is satisfied. The deferred action is applied by the *eval* operation.

The induction step cases correspond to the statements, snapshot2 $x$ $\{s2\}$, let $x = e$ in $s2$, let $x = \text{create}(c)$ in $s2$, and $s2; s2$, and can be proved directly from the induction hypothesis as they do not perform actions on the model by themselves. □

PROOF. Theorem 7. By induction on the structure of first-level statements. The base cases correspond to:

- create($c$): rule (E-CreateRoot) satisfies the theorem since references are not created.
- delete($x$): rule (E-DeleteRoot) satisfies the theorem since the deletion of the object referenced by $x$ does not create any dangling reference as it has to be isolated.
- snapshot $x$ $\{s2\}$, we use Lemma 8 to show that rule (E-Snapshot) generates a valid model.
- () is in canonical form and satisfies the theorem vacuously.

The induction step corresponds to the statements let $x = e$ in $s$, let $x = \text{create}(c)$ in $s$ and $s; s$, which can be directly proved from the induction hypothesis given that they do not apply actions to the model. □

# E. TYPE SYSTEM

## E.1 Type system

We provide a type system for the transformation language that augments the set of typing rules presented in Table 1 and in Table 2.

Three kinds of typing judgements are introduced, $\Gamma \mid \Pi \vdash s : \tau$ for first-level statements, $c \mid \Gamma \mid \Pi \vdash s : \tau$ for second-level statements, and $\Gamma \mid \Pi \vdash e : \alpha$ for expressions. In typing judgements for second-level statements, $c$ denotes the class name of the object under focus.

Table 13 gathers the typing relation for expressions, borrowing two rules from Table 1, Table 14 gathers the typing relation for first-level statements and Table 14 gathers the typing relation for second-level statements.

## E.2 Consistency Theorem

The following theorem states that evaluation relation for FMA programs preserves the type of the models being transformed.

*Theorem 8.* Consistency for $\Downarrow_s$. Given a structured model $M = (os, \Pi)$, and a configuration $k$ with $M$ and a well-typed, closed FMA statement $s$, if $M : \mathcal{M}$ and $k \Downarrow_s k'$ then, for some $\Pi' \supseteq \Pi$, $M' = (k'_{os}, \Pi')$ is a structured model and $M' : \mathcal{M}$.

Showing the soundness of the typing rule for the snapshot statement requires proving that replacing a modified object in the model is consistent, which can be done following a strategy similar to the one followed in the proof of the validity theorem for first-level statements.

The following lemma guarantees that augmenting the typing environment for objects identifiers preserves the typing of the objects already present in the model.

*Lemma 9.* Weakening lemma for reference typing. Given $M = (os, \Pi)$, if $M : \mathcal{M}$, and for some $\Pi' \supseteq \Pi$, then $(os, \Pi') : \mathcal{M}$.

PROOF. By induction on typing derivations. □

We need an additional lemma to prove the soundness of the type system for first-level statements with respect to $\Downarrow_s$. The object replacement lemma guarantees that plugging back a well-typed object into a set of objects obtained by unplugging that object, preserves the consistency of the model.

*Lemma 10.* Object replacement. Given $M = (os, \Pi)$ such that $M : \mathcal{M}$, a configuration $k$ with $M$, where $k = \eta \mid os \mid s$ for a closed FMA program $s$, and a second-level statement $s2$ closed with respect to $\eta$, if $< os'; \rho > = os|_l$ and

$$\eta \mid os' \mid l \mid \rho_{ps} \mid \emptyset \mid s2 \Downarrow_{s2} \eta' \mid os' \mid l'' \mid ps' \mid as \mid \bullet$$

then, for some $\Pi' \supseteq \Pi$, $(eval(as, \eta', <\rho_o \mid ps'>[os']_l), \Pi') : \mathcal{M}$.

PROOF. In Lemma 1, we already showed that the evaluation of second-level statements, together with the *eval* operation, preserve the validity of models. We will focus on showing that the evaluation of second-level statements preserves metamodel conformance.

By induction on the typing relation on $s2$, we proceed by case analysis on the final typing rule used. The base cases are:

- T-Create: we can only apply the rule E-Create, producing $\eta' \mid os \mid l \mid p = os'' \rho', ps \mid as \mid \bullet$.

  Lemma 9 ensures that augmenting the typing environment $\Pi$ with a new object identifier preserves the typing of model $M$.

  By definition $\Gamma \mid \Pi' \vdash \rho' : c'$ and, by assumption, the new object is a value of the type of the property or of its subtypes, i.e. $c \mid \Gamma \mid \Pi \vdash p : \text{ref } c' <: \mathcal{M}_{cl}(c)|_p$. By rules (T-Obj1) and (T-Obj2), (T-Prop) and (T-Obj), we have $\Gamma \mid \Pi \vdash (p = os'' \rho', ps) : \mathcal{M}_{cl}(c)$. This means that the *eval* operator replaces the object at location $l'$ with an object with the same object type.

- T-AttSet: we can only apply the rule E-AttSet and we obtain $\eta \mid os \mid l \mid p = v', ps \mid as \mid \bullet$.

| error | representation | operation (contracts) |
|---|---|---|
| > 1 container | ✓ | — |
| no containment cycles | ✓ | — |
| no parallel edges (optional) | ✓ ordered sets | E-REFUNISET, E-REFBISET, E-CMTUNISET and E-CMTBISET: ensured by postconditions |
| dangling references | — | E-REFBISET and E-CMTBISET: prevented by preconditions E-REFBISET, E-CMTBISET, E-REFBIUNSET and E-CMTUNSET: ensured by postconditions |
| inconsistent opposite references | — | E-REFBISET, E-REFBIUNSET, E-CMTBISET and E-CMTUNSET: ensured by postconditions |

Table 12: Summary of trapped errors

$$\frac{a \in D_b}{\Gamma \mid \Pi \vdash a : b} \quad \text{(T-BASE)}$$

$$\frac{\Pi(o) = c}{\Gamma \mid \Pi \vdash a : ref\ c} \quad \text{(T-REF)}$$

$$\frac{x : \alpha \in \Gamma}{\Gamma \mid \Pi \vdash x : \alpha} \quad \text{(T-VAR)}$$

Table 13: Typing rules for expressions.

$$\frac{notAbstract(c) \quad \mathcal{M}_{cl}(c) <: \mathcal{M}_{cl}(\mathcal{M}_r)}{\Gamma \mid \Pi \vdash \texttt{create}(c) : ()} \quad \text{(T-CREATEROOT)}$$

$$\frac{x : ref\ c \in \Gamma \quad \mathcal{M}_{cl}(c) <: \mathcal{M}_{cl}(\mathcal{M}_r)}{\Gamma \mid \Pi \vdash \texttt{delete}(x) : ()} \quad \text{(T-DELETEROOT)}$$

$$\frac{\Gamma \mid \Pi \vdash e : \alpha \quad \Gamma, x : \alpha \mid \Pi \vdash s : \tau}{\Gamma \mid \Pi \vdash \texttt{let}\ x = e\ \texttt{in}\ s : \tau} \quad \text{(T-FMALET)}$$

$$\frac{\Gamma \mid \Pi \vdash \texttt{create}(c) : () \quad \Gamma, x : ref\ c' \mid \Pi \vdash s : \tau}{\Gamma \mid \Pi \vdash \texttt{let}\ x = \texttt{create}(c)\ \texttt{in}\ s : \tau} \quad \text{(T-FMALETCREATE)}$$

$$\frac{\Gamma \mid \Pi \vdash x : ref\ c \quad c \mid \Gamma \mid \Pi \vdash s2 : ()}{\Gamma \mid \Pi \vdash \texttt{snapshot}\ x\ \{s2\} : ()} \quad \text{(T-FMASNAPSHOT)}$$

$$\Gamma \mid \Pi \vdash () : () \quad \text{(T-FMAUNIT)}$$

$$\frac{\Gamma \mid \Pi \vdash s_1 : () \quad \Gamma \mid \Pi \vdash s_2 : \tau}{\Gamma \mid \Pi \vdash s_1\ ;\ s_2 : \tau} \quad \text{(T-FMASEQ)}$$

Table 14: Typing rules for first-level statements.

$$\frac{notAbstract(c') \quad c \mid \Gamma \mid \Pi \vdash (p : c') <: \mathcal{M}_{cl}(c)|_p}{c \mid \Gamma \mid \Pi \vdash \texttt{create}(p, c') : \bullet} \quad \text{(T-CREATE)}$$

$$\frac{\Gamma \mid \Pi \vdash e : \alpha \quad c \mid \Gamma \mid \Pi \vdash (p : \alpha) <: \mathcal{M}_{cl}(c)|_p}{c \mid \Gamma \mid \Pi \vdash \texttt{set}(p, e) : \bullet} \quad \text{(T-SET)}$$

$$\frac{c \mid \Gamma \mid \Pi \vdash (p : b) := \mathcal{M}_{cl}(c)|_p}{c \mid \Gamma \mid \Pi \vdash \texttt{unset}(p) : \bullet} \quad \text{(T-ATTUNSET)}$$

$$\frac{x : ref\ c' \in \Gamma \quad c \mid \Gamma \mid \Pi \vdash (p : ref\ c') <: \mathcal{M}_{cl}(c)|_p}{c \mid \Gamma \mid \Pi \vdash \texttt{unset}(p, x) : \bullet} \quad \text{(T-REFUNSET)}$$

$$\frac{x : ref\ c' \in \Gamma \quad c \mid \Gamma \mid \Pi \vdash p : c' <: \mathcal{M}_{cl}(c)|_p}{c \mid \Gamma \mid \Pi \vdash \texttt{set}(p, x) : \bullet} \quad \text{(T-CMTSET)}$$

$$\frac{x : ref\ c' \in \Gamma \quad c \mid \Gamma \mid \Pi \vdash p : c' <: \mathcal{M}_{cl}(c)|_p}{c \mid \Gamma \mid \Pi \vdash \texttt{unset}(p, x) : \bullet} \quad \text{(T-CMTUNSET)}$$

$$\frac{\Gamma \vdash e : \alpha \quad c \mid \Gamma, x : \alpha \mid \Pi \vdash s2 : \tau}{c \mid \Gamma \mid \Pi \vdash \texttt{let}\ x = e\ \texttt{in}\ s2 : \tau} \quad \text{(T-ACTLET)}$$

$$\frac{c \mid \Gamma \mid \Pi \vdash \texttt{create}(p, c') : \bullet \quad c \mid \Gamma, x : ref\ c' \mid \Pi \vdash s2 : \tau}{c \mid \Gamma \mid \Pi \vdash \texttt{let}\ x = \texttt{create}(p, c')\ \texttt{in}\ s2 : \tau} \quad \text{(T-ACTLETCREATE)}$$

$$\frac{\Gamma \mid \Pi \vdash x : ref\ c \quad c \mid \Gamma \mid \Pi \vdash s2 : ()}{\Gamma \mid \Pi \vdash \texttt{snapshot2}\ x\ \{s2\} : ()} \quad \text{(T-ACTSNAPSHOT)}$$

$$c \mid \Gamma \mid \Pi \vdash \bullet : \bullet \quad \text{(T-ACTUNIT)}$$

$$\frac{c \mid \Gamma \mid \Pi \vdash s2_1 : \bullet \quad c \mid \Gamma \mid \Pi \vdash s2_2 : \tau}{c \mid \Gamma \mid \Pi \vdash s2_1\ ;\ s2_2 : \tau} \quad \text{(T-ACTSEQ)}$$

Table 15: Typing rules for second-level statements.

- By definition $\Gamma \vdash v' : b$ and, by assumption, $b$ is the type of the property, i.e. $p : ref\ b \in \mathcal{M}_{cl}(c)$. By rules (T-Prop) and (T-Obj), we have
  $\Gamma \mid \Pi \vdash (p = v', ps) : \mathcal{M}_{cl}(c)$. This means that the *eval* operator replaces the object at location $l'$ with an object with the same object type.

- T-AttUnset: similar reasoning as above, but the new value is the default value of the data type, which is guaranteed to be a well-typed value for the property.

- T-RefSet: we can apply to evaluation rules
  - E-RefUniSet, producing
    $$\eta \mid os \mid l \mid p = (o \in is)?\ is : is\ o, ps \mid as \mid \bullet$$
    By assumption, $c \mid \Gamma \mid \Pi \vdash p : ref\ c' <: \mathcal{M}_{cl}(c)\mid_p$ and $\Gamma \mid \Pi \vdash x : ref\ c'$. When the object identifier is already in the collection, the theorem is trivially proved for this case. Otherwise, by rules T-Ref, T-Ref1, T-Ref2 and T-Prop, we have $\Gamma \mid \Pi \vdash (p = is\ o, ps) : \mathcal{M}_{cl}(c)$. The conclusion implies that the *eval* operator replaces the object at location $l'$ with an object with the same object type.
  - E-RefBiSet, producing
    $$\eta \mid os \mid l \mid p = (o \in is)?\ is : is\ o \mid as\ set(o, p', last(l)) \mid \bullet$$
    The reasoning for this case is very similar but in addition, we have to show that the evaluation of deferred action preserves conformance. The semantics of a deferred set action on the model is performed by the *eval* operation and corresponds to E-RefUniSet. The proof follows a similar argument as in the unidirectional case.
  - E-CmtUniSet The reasoning is similar to the one for E-RefUniSet.
  - E-CmtBiSet. The reasoning is similar to the one for E-RefBiSet.

- T-RefUnset: we can apply two rules: E-RefUniUnset and E-RefBiUnset. The reasoning is similar to the one for T-RefSet.

- T-CmtUnset: we can apply the rule E-CmtUnset and the reasoning is similar to the one for T-RefSet.

- T-ActUnit: $\bullet$ is in normal form by Lemma 6 and it vacuously satisfies the condition in the theorem since we cannot apply any rule.

The induction step cases correspond to the typing rules T-ActSnapshot, T-ActLet, T-ActLetCreate, T-ActSeq, which can be proved directly from the induction hypothesis, as they do not perform any changes on the model by themselves.  □

Now we are ready to prove the consistency theorem:

Proof. (Theorem 8) By induction on a typing derivation with respect to the typing relation for first-level statements. At each step of the induction, we assume the desired property holds for all subderivations and proceed by case analysis on the final rule of the derivation.

We will assume that $M = (os, \Pi)$ and that $\emptyset; \Pi \vdash os : c'$.

- T-CreateRoot: the only evaluation rule that can be applied is (E-CreateRoot) and we obtain $\eta' \mid os\ \rho' \mid ()$ where $\Gamma \mid \Pi' \vdash \rho' : c$ by definition with $\Pi' = \Pi \cup \{\rho'_o \mapsto c\}$. By Lemma 9, $\Gamma \mid \Pi' \vdash os : c'$. We have that:
  - $cl(c) <: cl(c')$ and by rule (T-Obj1), $\Gamma \mid \Pi' \vdash os\ \rho' : c'$;
  - $cl(c') <: cl(c)$ and by rule (T-Obj2), $\Gamma \mid \Pi' \vdash os\ \rho' : c$.

  By rule (T-Prop) and by Definition 6, $(os\ \rho', \Pi') : \mathcal{M}$.

- T-DeleteRoot: we can only apply rule (E-DeleteRoot), which deletes a root object producing the configuration $\eta \mid os \mid ()$. The remaining collection of objects $os$ is either a well-typed collection of objects or the empty model $\emptyset$. In the second case, the empty set of objects is a valid value of any object type by rule T-ObjAny and by the rules that define the object subtyping relation in Definition 4.

- T-FmaSnapshot: the only evaluation rule that we can apply is (E-Snapshot), which means that we must have $\eta \vdash x \Downarrow_e o,\ l' = \eta(o),\ \langle os', \rho'' \rangle = os\mid_{l'}$ and
  $$\eta \mid os' \mid l' \mid \rho''_{ps} \mid \emptyset \mid s2 \Downarrow_{s2} \eta' \mid os'' \mid l'' \mid ps' \mid as \mid \bullet$$
  By Lemma 10, we have that, for some $\Pi' \supseteq \Pi$ such that $\Pi'$ specifies the typing for the fresh object identifiers that may have been obtained by the rule E-Create if any create statement is in $s2$,
  $$(eval(as, \eta', \texttt{<}\rho_o \mid ps'\texttt{>}[os'']_l), \Pi') : \mathcal{M}$$
  is a structured model.

The induction step cases correspond to T-FmaLet, T-FmaLetCreate, T-FmaSeq, which can be proved directly from the induction hypothesis as the rules that can be applied do not apply changes to the model by themselves.  □